\documentclass[12pt]{article}
\usepackage{latexsym}
\usepackage{amsmath}
\usepackage{amssymb}
\usepackage{relsize}
\usepackage{geometry}

\numberwithin{equation}{section}

\usepackage{tensor}
\usepackage{physics} 
\usepackage{csquotes} 

\usepackage{graphicx}       	
\usepackage[font=small,labelfont=bf]{caption}
\usepackage{subcaption}

\def\beqn{\begin{eqnarray}}
\def\eeqn{\end{eqnarray}}

\def\beq{\begin{equation}}
\def\eeq{\end{equation}}
\def\ba{\beq\new\begin{array}{c}}
\def\ea{\end{array}\eeq}

\def\Tr{{\rm Tr}}
\newcommand{\gsim}{\lower.7ex\hbox{$
\;\stackrel{\textstyle>}{\sim}\;$}}
\newcommand{\lsim}{\lower.7ex\hbox{$
\;\stackrel{\textstyle<}{\sim}\;$}}

\newcommand{\ntwo}{${\mathcal N}=2$ }

\newcommand{\ntwot}{${\mathcal N}= \left(2,2\right) $ }

\newcommand{\none}{${\mathcal N}=1$ }

\newcommand{\cpn}{$\mathbb{CP}(N-1)\;$}

\newcommand{\wcpNt}{$\mathbb{WCP}(N,N_f-N)\;$}

\def\slashed#1{\setbox0=\hbox{$#1$}             
   \dimen0=\wd0                                 
   \setbox1=\hbox{/} \dimen1=\wd1               
   \ifdim\dimen0>\dimen1                        
      \rlap{\hbox to \dimen0{\hfil/\hfil}}      
      #1                                        
   \else                                        
      \rlap{\hbox to \dimen1{\hfil$#1$\hfil}}   
      /                                         
   \fi}                                        %

\usepackage{mathtools}
\usepackage{hyperref}
\usepackage[table]{xcolor}
\usepackage{tikz-cd}
\usepackage{placeins}
\usepackage[toc,page, title, titletoc]{appendix}
\newcommand{\mathsym}[1]{{}}
\newcommand{\unicode}[1]{{}}

\begin{document}

\hypersetup{%
linkbordercolor=blue,
}

%
%

\begin{titlepage}

\begin{flushright}
FTPI-MINN-26-15 \\ UMN-TH-4535/26
\end{flushright} 

\begin{center}

{\Large{\bf 
Unified description of  \\ color-electric and  color-magnetic strings \\
\vspace{3mm}  in hybrid vacua  of ${\mathcal N}=2$ supersymmetric QCD
}}

\vspace{5mm}

{\large  \bf E.~Ievlev$^{\,a}$, D.~Vasilev$^{\,b,c,d,e}$ and  A.~Yung$^{\,b,f}$}

\end{center}

\begin{center}

{\it  $^{a}$William I. Fine Theoretical Physics Institute,
University of Minnesota,
Minneapolis, MN 55455, USA}\\

$^{b}${\it NRC ``Kurchatov Institute'' - PNPI,  Gatchina, St. Petersburg
				188300, Russia}\\

$^{c}${\it  St. Petersburg State University, Universitetskaya nab., St.~Petersburg \\ 199034, Russia}\\

{\it $^d$ Steklov Mathematical Institute, Fontanka 27, St. Petersburg 
 191023, Russia}\\
{\it $^e$ ITMO University, St.~Petersburg,
 197101, Russia}\\
{\it $^f$ HSE University, St. Petersburg,
				194100, Russia}\\

\end{center}

\vspace{5mm}

\begin{center}
{\large\bf Abstract}
\end{center}

We study non-Abelian strings supported in $r$-vacua of 4D $\mathcal{N}=2$ supersymmetric QCD (SQCD) with gauge group $U(N)$ and $N_f$ quark hypermultiplets ($N_f\ge N$), where $r<N$ (scalar) quarks develop vacuum expectation values. At the quantum level, $N-r-1$ monopoles in the orthogonal sector of the gauge group also form a condensate. This leads to the formation of flux tubes (strings) that carry both color-electric and color-magnetic fluxes and confine quarks and monopoles, respectively. We present a unified description of the 2D effective theory living on the world sheet of this non-Abelian string. In particular, we derive the exact twisted superpotential of the world sheet theory, which is deformed by the presence of non-perturbative gaugino condensate in 4D hybrid vacua, using the Gaiotto–Gukov–Seiberg method of resolvents. Next, we extrapolate the 2D-4D correspondence of BPS spectra of the world sheet theory and 4D SQCD (known for quark vacua) to hybrid vacua. Namely, our key result is the precise matching between the masses of the two-dimensional kinks and the four-dimensional confined monopoles or quarks. In addition, we provide a physical picture of the quark/monopole confinement in hybrid vacua.

\end{titlepage}

\setcounter{tocdepth}{2}
\numberwithin{equation}{section}
\tableofcontents


\section {Introduction }
\label{intro}

The search for non-Abelian generalizations of the Seiberg-Witten scenario of confinement \cite{SW,SW2} led to the discovery of non-Abelian vortex strings (flux tubes) \cite{HT1,ABEKY,SYmon,HT2} in four-dimensional (4D) $\mathcal{N} =2$ supersymmetric QCD (SQCD)  (see \cite{Trev,Jrev,SYrev,Trev2} for reviews).
In the basic setup with gauge group $U(N)$ and $N_f\ge N$ quark hypermultiplets, the Fayet-Iliopoulos deformation \cite{FI} triggers squark condensation, which leads to the formation of  non-Abelian  strings  with internal orientational and size moduli.

While the squark vacuum expectation values (VEV)  break both the gauge and the flavor symmetries, a diagonal subgroup $SU(N)_{C+F}$ survives  (color-flavor locking).
This global symmetry gives rise to additional zero modes on the vortex, giving rise to rich world-sheet dynamics.
In the case with $N_f=N$, the vortex world sheet supports the $\mathbb{CP}^{N-1}$ sigma model, while for $N_f>N$ the semilocal sector is described by its weighted-projective generalizations 
\wcpNt  \cite{HT1,HT2,Vachaspati:1991dz,AchVas,SYsem,Jsem,SVY}.

The world-sheet theory is a helpful tool for studying four-dimensional physics.
In the Higgs phase of 4D SQCD, confined monopoles appear as junctions of distinct elementary strings, and in the two-dimensional description they are realized as kinks interpolating between different  vacua of the world-sheet theory \cite{SYmon}. 
This identification underlies the 2D-4D correspondence: protected quantities in the non-Abelian string world-sheet theory, most notably,  the Bogomol'nyi-Prasad-Sommerfield (BPS) spectrum of states in  the 2D world-sheet theory, coincide with the BPS spectrum of 4D states in the quark vacuum  given by the exact Seiberg-Witten solution \cite{SW2} on the Coulomb branch. This coincidence was observed in \cite{Dorey,DoHoTo} in quark  vacua and explained
later in \cite{SYmon,HT2}  using the picture of  4D monopoles confined by the non-Abelian string, which are seen as kinks in the 2D world-sheet theory.

Upon introducing the deformation of $\mathcal N=2$ SQCD by a small mass $\mu$ of the adjoint matter, the Coulomb branch is mostly lifted, leaving a family of the so-called $r$ vacua. 
In the semiclassical regime, the integer $r$ counts the number of condensed quark flavors and ranges from $0$ to $N$. 
Earlier work established that the physics depends sharply on which $r$ vacuum is chosen. 
In the fully Higgsed $r=N$ quark vacuum, the world-sheet description of the non-Abelian string can be formulated in terms of the  weighted $\mathbb{CP}$  model, denoted as \wcpNt , see, for example, the review \cite{SYrev}.

The case of the $r=N-1$ vacuum was studied in \cite{Shifman:2014lba}. It turns out that the non-perturbative gaugino condensate absent in the $r=N$ vacuum feeds directly into the string world-sheet dynamics and deforms the effective description. However, the 2D-4D BPS correspondence, i.e.   the matching of spectra of 2D kinks and 4D  confined monopoles, remains intact 
\cite{Shifman:2014lba}.

A natural generalization addressed in this paper is to formulate a unified treatment of the non-Abelian string world-sheet theory throughout the entire interval
\begin{equation}
	0\le r\le N,
\end{equation}
and to determine how the known special cases fit into a single framework. Physically, an important new
feature of  hybrid  $r<N-1$ vacua is the presence of $(N-r-1)$ condensed monopoles  in addition to $r$ condensed quarks \footnote{Condensed quarks and monopoles belong to orthogonal sectors of the gauge group and therefore, are mutually local, see below.} .  Due to this condensation, both color-magnetic and color-electric strings are formed, which confine monopoles and quarks, respectively. We present a unified description of  the 2D effective theory living on the world sheet of this confining non-Abelian  string. In particular, we derive the exact twisted superpotential of the world sheet theory using
the Gaiotto–Gukov–Seiberg method of resolvents \cite{Gaiotto:2013sma}. 

In particular, this means that Abrikosov-Nielsen-Olesen Abelian strings \cite{ANO} formed due to the monopole condensation and   responsible for the confinement of quarks in the Seiberg-Witten scenario are seen as particular vacua in our world-sheet theory on the non-Abelian string.

We also show that the 2D-4D correspondence is still valid in hybrid vacua. We calculate BPS kink masses in the world-sheet theory on the non-Abelian string using the exact superpotential and compare them with the masses of monopoles and quarks on the Coulomb branch  at $\mu\to 0$ given by the exact Seiberg-Witten solution \cite{SW,SW2}. We find an exact match.

The paper is organized as follows. In Sec.~\ref{sec:sqcd} we briefly review \ntwo SQCD, its $r$ vacua, and the Seiberg-Witten curve. In Sec.~\ref{subsec:non-ab_strings} we review the world sheet theory on the non-Abelian string in the $r=N$ vacuum and discuss the classical world sheet theory for the non-Abelian  string 
in $r<N$ vacua. In Sec.~\ref{sec:confiniment} we review the classical picture of confinement of monopoles and quarks in hybrid vacua; in particular, we consider the simplest case of the $r=1$ vacuum in $U(3)$ SQCD. In Sec.~\ref{sec:Quantum_world_sheet} we derive the exact quantum superpotential 
in the world sheet theory, and in Sec.~\ref{2D_4D_corr} calculate kink masses and demonstrate the 2D-4D correspondence in hybrid $r$ vacua. Sec.~\ref{sec:concl} contains our conclusions, while Appendices present some further examples and the details of our calculations.

\section {\ntwo supersymmetric QCD in 4D and its $r$-vacua }
\label{sec:sqcd}

In this section, we briefly review our setup for the 4D $ \mathcal{N}=2 $ SQCD with a massive deformation of the adjoint matter. 
We discuss the structure of the $r$-vacua and the corresponding Seiberg-Witten curve.

\subsection{The 4D setup and quasiclassical analysis }

We start with a 4D gauge theory with the gauge group $U(N) \equiv (SU(N) \cross U(1))/Z_N$.
The $\mathcal{N}=2$ vector multiplet consists of the gauge field itself (vector), two gauginos (Weyl fermions), and a complex scalar, which we denote as $(a,\lambda^1,\lambda^2,F_{\mathcal{A}})$ for the $U(1)$ part and $(a^a,\lambda^{1a},\lambda^{2a},F^a)$ for the $SU(N)$ sector, respectively. 
Here, $a=1,\ldots,N^2-1$.
Each pair of gauginos $(\lambda^1,\lambda^2)$, $(\lambda^{1a},\lambda^{2a})$ forms a doublet under the global $SU(2)_R$ symmetry. 
The matter sector is represented by the hypermultiplets $(\psi^{kA},q^{kA}, \bar{\tilde{\psi}}^{kA},\bar{\tilde{q}}^{kA})$, all in the fundamental representation of $SU(N)$. Here, $k=1,\ldots,N$ is the color index, and  $A=1,\ldots,N_f$ is the flavor index.
The scalar components $(q^{kA},\bar{\tilde{q}}^{kA})$ (squarks) also form an $SU(2)_R$ doublet. 
In this work we focus on the case
\begin{align}
	N \le N_f < 2N,
\end{align} 
which ensures asymptotic freedom. 
The \ntwo  part of the superpotential is 
\begin{align}
	\mathcal{W}_{\mathcal{N}=2}= \sum_{A=1}^{N_f}\left\{\sqrt{2} \, \tilde{Q}_A \left(\frac{1}{2}\mathcal{A}+\mathcal{A}^a T^a \right) Q^A + m_A \tilde{Q}_A Q^A \right\},
\label{N=2_superpotential_4D}
\end{align}
where $\mathcal{A}$,  $\mathcal{A}^a$ are \none chiral superfields with lowest components $a$, $a^a$, and  $Q^{kA}$, $\tilde{Q}_{Ak}$ are chiral quark superfields, while $m_A$ denotes the quark masses; see the review \cite{SYrev} for details.

In order to trigger squark condensation, we, following \cite{SW,SW2}, introduce a small mass term for the adjoint matter 
\begin{align}
	\mathcal{W}_{\mathrm{def}}=\frac{\mu}{2}\left(\frac{N}{2}\mathcal{A}^2+(\mathcal{A}^a)^2 \right). \label{def}
\end{align}
Strictly speaking, this deformation breaks supersymmetry down to $\mathcal{N}=1$, but  to leading order in  the limit $|\mu |/ \Lambda_{4D} \ll 1$ and if all quark
masses are equal, this    superpotential  reduces to the  Fayet--Iliopoulos $F$ term, which does not break \ntwo supersymmetry; see \cite{HanStrassZaf,VY}. Here, $ \Lambda_{4D}$ is the dynamical scale of SQCD.

In components, the bosonic part of the action is 
\begin{equation}
	\begin{split}
		S = \int d^4x \biggl[ \frac{1}{4g_2^2} (F_{\mu\nu}^a)^2 + \frac{1}{4g_1^2} (F_{\mu\nu})^2 + \frac{1}{g_2^2} |D_\mu a^a|^2 + \frac{1}{g_1^2} |\partial_\mu a|^2   \\ +|\nabla_\mu q^A|^2 + |\nabla_\mu \bar{\tilde{q}}^A|^2 + V(q^A, \tilde{q}_A, a^a, a) \biggl], \label{s}
	\end{split}
\end{equation}
$\nabla_{\mu}, D_{\mu}$  are covariant derivatives in the fundamental and in the adjoint representation, respectively,
\begin{align}
	\nabla_{\mu} = \partial_{\mu} - \frac{i}{2} A_{\mu} - i A_{\mu}^{a} T^{a}, \quad D_{\mu}=\partial_{\mu}-i A^a f^{abc} \label{cov}.
\end{align}
The scalar potential  is given by
\begin{align}
	V\left(q^A, \widetilde{q}_A, a^{a}, a\right)&= \nonumber \\
	&\frac{g_2^2}{2}\left(\frac{1}{g_2^2} f^{a b c} \bar{a}^b a^c+\bar{q}_A T^a q^A-\tilde{q}_A T^a \Bar{\tilde{q}}^A\right)^2 \nonumber \\
	& +\frac{g_1^2}{8}\left(\bar{q}_A q^A-\widetilde{q}_A \overline{\widetilde{q}}^A\right)^2\nonumber \\
	& +2 g_2^2\left|\widetilde{q}_A T^a q^A+\frac{\mu}{\sqrt{2}}a^a\right|^2+\frac{g_1^2}{2}\left|\widetilde{q}_A q^A+N\frac{\mu}{\sqrt{2}} a\right|^2 \nonumber \\
	& +\frac{1}{2} \sum_{A=1}^{N_{f}}\left\{\left|\left(a+\sqrt{2}m_A+2a^a T^a\right) q^A\right|^2\right. + \left.\left|\left(a+\sqrt{2}m_A+2a^a T^a\right) \Bar{\tilde{q}}^A\right|^2\right\}. 
	\label{V}
\end{align}
Here,
$g_1$ and $ g_2$ are $U(1)$ and $SU(N)$ coupling constants, and we employ the standard conventions for the structure constants $f^{abc}$ and the generator normalization 
\begin{align}
	\mathrm{Tr}(T^aT^b)=\frac12\delta^{ab}.
\label{algebra_normalization}
\end{align}

\begin{table}[t]
	\centering
	\begin{tabular}{|c|c|c|}
		\hline
		\textbf{Energy scale} & \textbf{Gauge symmetry} & \textbf{Flavor-related symmetry} \\
		\hline
		$m \ll E$  &  $U(N)$  & $SU(N_f)_\text{flavor}$ \\
		\hline \rowcolor{gray!15} 
		$E \sim m$ & \multicolumn{2}{|c|}{ Adjoint scalar condensation (classical), Eq.~\eqref{phi}  } \\
		\hline
		$\sqrt{\xi} \sim \sqrt{\mu m} \ll E \ll m$  &  $U(r) \times  U(N-r)$  & $SU(N_f)_\text{flavor}$ \\
		\hline \rowcolor{gray!15} 
		$E \sim \sqrt{\xi}$ & \multicolumn{2}{|c|}{Squark condensation, Eq.~\eqref{q} } \\
		\hline
		$\Delta m \ll E \ll \sqrt{\xi}$  &  $U(N-r)$  & $ SU(r)_\text{c+f,lock} \times SU(N_f-r)_\text{flavor} \times U(1)$   \\
		\hline \rowcolor{gray!15} 
		$E \sim \Delta m$ & \multicolumn{2}{|c|}{Flavor symmetry breaking}  \\
		\hline
		$\Lambda \ll E \ll \Delta m$  & $U(N-r)$  & $U(1)^{N_f-1}$ \\ 
		\hline \rowcolor{gray!15} 
		$E \sim \Lambda$ & \multicolumn{2}{|c|}{ Adjoint scalar condensation (non-pert) }  \\
		\hline
		$\sqrt{\mu \Lambda} \ll E \ll \Lambda$  & $U(1)^{N-r}$  & $U(1)^{N_f-1}$ \\
		\hline \rowcolor{gray!15} 
		$E \sim \sqrt{\mu \Lambda}$ & \multicolumn{2}{|c|}{ Monopole condensation }  \\
		\hline
		$E \ll \sqrt{\mu \Lambda}$ & $U(1)_\text{unbr}$ & $U(1)^{N_f-1}$ \\
		\hline
	\end{tabular}
	\caption{Scale hierarchy for generic $r$  }
	\label{tab:scales_hierarchy_table_normal}
\end{table}

As one can see, there are a number of various scales in this theory.
For now, let us consider the hierarchy outlined in Table~\ref{tab:scales_hierarchy_table_normal}.
Let us describe the $r$-vacua that we are interested in.
For an integer $r \leqslant N$, we consider a scenario where the adjoint scalars classically develop  VEVs,
\begin{align}
	\Phi_{\mathrm{cl}} = \bigg<{\frac{1}{2}a+a^a T^a}\bigg>_{\mathrm{cl}}= -\frac{1}{\sqrt{2}}\begin{pmatrix}
		m_1 & 0& \ldots& \ldots & \ldots&0 \\
		0 & \ddots& 0 & \ldots & \ldots &0 \\
		\vdots& 0& m_{r}& 0 &\ldots& 0 \\
		\vdots & \vdots & 0 & 0 & \ldots & 0 &  \\
		\vdots & \vdots & \vdots & \vdots& \ddots & \vdots \\
		0 & 0& 0 & 0& \ldots & 0 
	\end{pmatrix}, 
\label{phi}
\end{align}
Explicitly, the component VEVs are given by
\begin{align}
	\big< a\big>_{\mathrm{cl}} = -\frac{\sqrt{2}}{N} \sum_{A=1}^{r}m_A, \ \big< a^a\big>_{\mathrm{cl}}= 2  \mathrm{Tr}\left( T^{a} \Phi_{\mathrm{cl}}\right) \,.
\label{a}
\end{align}
In this case, at the onset of the adjoint VEV, the gauge group is broken down to
\begin{align}
	U(N)  \overset{|m_P|}{\xrightarrow{\hspace{1cm}}} U(r) \cross U(N-r) \,.
\label{gaugebreak}
\end{align}
At a lower scale, $r$ of the scalar quarks also develop VEVs
\begin{align}
	\ \big< \Bar{\tilde{q}}^{jA}\big>_{\mathrm{cl}} = \big< q^{jA }\big>_{\mathrm{cl}} = \frac{1}{\sqrt{2}}\begin{pmatrix}
		\sqrt{\xi^{\mathrm{cl}}_1 } & 0& \ldots& \ldots &\ldots  & \ldots &\ldots & 0 \\
		0 & \ddots& 0 & \ldots & \ldots& \ldots & \ldots & 0  \\
		\vdots& 0& \sqrt{\xi^{\mathrm{cl}}_{r}}& 0 &\ldots&  \ldots& \ldots & 0\\
		\vdots & \vdots & 0 & 0 & \ldots & 0 & \ldots & 0\\
		\vdots & \vdots & \vdots &\vdots & \ddots & \vdots &\ddots & \vdots \\
		0 & 0 & 0 & 0& \ldots & 0 & \ldots &0 
	\end{pmatrix},  
	\label{q}
\end{align}
where $\xi^{\mathrm{cl}}_P=2 \mu m_P$ follows from the potential \eqref{V}.
The quark masses in this vacuum in the limit $\mu\to 0$ are given by
\begin{align}
	m^{q}_{k,A} = |m_A-m_k|, \ k=1,\ldots, r \ ;\ m^{q}_{r+1,A}= \ldots =m^{q}_{N,A}=|m_A|. \label{q_masses}
\end{align}

The VEV \eqref{q} further breaks the gauge group \eqref{gaugebreak} down to $U(N-r)$. 
The broken $U(r)$ part of the gauge group combines with the broken $U(r) \subset SU(N_f)$ part of the flavor symmetry to form a color-flavor-locked global symmetry 
\begin{equation}
	U(r)_\text{gauge} \times SU(r)_\text{flavor} \to SU(r)_\text{C+F} \times SU(N_f-r)\times U(1) \,.
	\label{c+f}
\end{equation}
This global symmetry is responsible for non-Abelian moduli living on the world sheet of the confining string; see Sec.~\ref{subsec:non-ab_strings}.

If the bare quark masses $m_P$ in the Lagrangian are not equal to each other, the global group $SU(r)_\text{C+F}$ is further broken below the scale $|\Delta m_{PK}|$.

The  remaining $U(N-r)$ part is broken at the quantum level  
through the Seiberg–Witten
mechanism, to which we now turn our attention.

\subsection{Seiberg-Witten solution}

To take into account the quantum effects, we go to the limit of small  $\mu$, where 
the theory is close to  $\mathcal{N}=2$ SQCD; this allows us to use the exact Seiberg-Witten (SW) solution \cite{SW,SW2}. 
The SW curve in our case reads \cite{Argyres:1996eh,HaOz}, 
\begin{align}
	y^2=\prod_{P=1}^N\left(x-\phi_P\right)^2-4\left(\frac{\Lambda_{4D}}{\sqrt{2}}\right)^{2 N-N_f} \prod_{A=1}^{N_f}\left(x+\frac{m_A}{\sqrt{2}}\right) \label{swcurve},
\end{align}
where $\Phi=\text{diag}(\phi_1,\ldots,\phi_N)$. 
In the $r$-vacua with  $r<N$,   this curve factorizes \cite{Cachazo:2001jy}
\begin{align}
	y^2=\prod_{K=1}^{r}(x-e_K)^2\prod_{P=r+1}^{N-1}(x-e_P)^2(x-e_{N}^{-})(x-e_{N}^{+}),
	\label{SW_curve_rep-01}
\end{align}
In the representation \eqref{SW_curve_rep-01}, the first $r$ double roots correspond to the condensation of the squarks, while the last $N-r-1$ are directly related to the condensed monopoles or dyons.
It is shown in  \cite{Cachazo:2001jy,Balasubramanian:2003tv} that the sum of $e_{N}^{+}$ and $e_{N}^{-}$ vanishes for the specific deformation \eqref{def}
\begin{align}
	e_{N}^{+}+e_{N}^{-}=0,
\end{align}
Furthermore, these two roots also determine the gluino condensate
\begin{align}
	\frac{\left< \Tr{W_{\alpha}W^{\alpha}}\right>}{32 \pi^2 \mu }\equiv\frac{S}{\mu} = \frac{(e_N^{\pm})^2}{2}. \label{glu_cond}
\end{align}
Note that the quantity \( S / \mu \) is finite in the limit \( \mu \to 0 \). Both the squark and monopole condensates  are determined by a common formula 
\cite{Shifman:2013zsa}.
\begin{align}
	\xi_{P}&=-2\sqrt{2}\mu \sqrt{(e_{P}-e_{N}^{+})(e_{P}-e_{N}^{-})}, \label{tension}
\end{align}
where one should choose $1 \le P\le r$ and $r+1 \le P \le N-1$ for squarks and monopoles, respectively. 

In the quasiclassical limit (large quark masses), the leading-order behavior of the roots can be read off Eq.~\eqref{phi},
\begin{align}
	&e_K \approx \phi_{K}\approx - \frac{m_K}{\sqrt{2}}, & \quad K=1,\ldots,r ;\\
	&e_P \approx \phi_P \approx 0, & P=r+1,\ldots N-1; \label{quasiroots}
\end{align}
In Appendix~\ref{sec:appendix_SW_technical}, we find quantum corrections to these values; we show that the parameters $\phi_{r+1},\ldots,\phi_{N}$ become non-vanishing. Their values are determined by the scale \( \Lambda_{4D} \) and break the remaining $U(N-r)$ sector \eqref{gaugebreak} down to $U(1)^{N-r}$. Then, as $N-r-1$ monopoles condense at the scale $\xi_{P} \sim|\mu \Lambda_{4D}|$, the gauge group is finally broken down to a single $U(1)_{\mathrm{unbr}}$.
This is the story for generic $r$; note, however, that $U(1)_\mathrm{unbr}$ doesn't survive in the $r=N$ case, which is a crucial distinction --- there are no long-range forces in that situation.  

Generically, for a given $r$ there are several isolated vacua.
One may pick $r$ condensed squarks out of the $N_f$ and then choose one of the $N-r$ monopole/dyon vacua, which yields 
\begin{equation}
	N_{r}= \frac{(N-r)N_f!}{r!(N_f-r)!}
\end{equation}
distinct vacua.
Our analysis below will concentrate on describing one representative case.

\section{Non-Abelian strings and their moduli space}
\label{subsec:non-ab_strings}

The condensation of electric and magnetic charges leads to the formation of strings that carry magnetic and electric  fluxes, respectively. In this section, we will provide a brief review of the non-Abelian magnetic strings  in the $r=N$ vacuum (see  \cite{HT1,ABEKY,SYmon,HT2} and \cite{SYrev} for a review) and, as a first step, introduce the classical effective  world-sheet model on the non-Abelian string in $r<N$ vacua, cf. \cite{Shifman:2014lba}.

\subsection{Recap of the $r=N$ case}
\label{sec:r=N}

First, we consider the case of $r=N=N_f$ and equal bare quark masses $\Delta m_{AB}\equiv m_A-m_B=0$. In this case, the gauge group breaks down completely at the squark VEV scale.  As 
\begin{align}
	\pi_1\left(SU(N) \cross U(1)/Z_{N}  \right) \neq 0,
\end{align}
$N$ Abelian strings (flux tubes) of $Z_N$ type arise in this vacuum. Each of them appears due to winding on a certain squark, e.g.
\begin{align}
	\phi_{\mathrm{string}} = \sqrt{2}\,\bar{\tilde{q}}^{kA} = \sqrt{2}\, q^{kA}\overset{|\vec{x}|\to \infty
	}{\to}\sqrt{\xi} \,\mathrm{diag}(1,\ldots,e^{i \alpha(\vec{x})}),
\end{align}
where $\vec{x}$ is a vector  in the $(x,y)$ plane orthogonal to the string, and $\alpha(\vec{x})$ is the polar angle. At the same time, the theory is in the color-flavor-locked phase ---  the global symmetry group is the diagonal part of $SU(N)_C \cross SU(N)_F $, see Eq.~\eqref{c+f}.
This global symmetry enables the strings to acquire additional zero modes.
Schematically, the transformation
\begin{align}
	\phi_{\text{Non-Abelian}}=U\phi_{\mathrm{string}}U^{-1}, \quad U\in SU(N)_C, U^{-1}\in SU(N)_F.
\end{align}
leaves the vacuum  intact, but brings a given string solution to another solution with the same energy.
Since $\phi_{\mathrm{string}}$ is stabilized by $SU(N-1)\cross U(1)$, the resulting moduli space is given by
\begin{align}
	SU(N)_{C+F} /( SU(N-1)\cross U(1)) \cong \mathbb{CP}(N-1),
\end{align} 
At $\mu\ll(m_A,\Lambda_{4D})$  (small deformation), the string is $1/2$ BPS saturated and conserves 4 supercharges.
The other $4$ SUSY generators act non-trivially and give rise to fermion zero modes. 

Now, relaxing the condition $\Delta m_{AB} =0 $ leads to the lifting of the moduli space, while adding ``extra'' quark flavors $N_f\ge N =r $ makes strings semilocal; their transverse size is not fixed,  and $2(N_f-N)$ new, so-called, size moduli appear on the world-sheet \cite{HT1,HT2,SYsem,Jsem,SVY}. 
The 2D effective world sheet theory describing the dynamics of internal moduli becomes a $\mathcal{N}=(2,2)$ twisted-mass-deformed $\mathbb{WCP}(N,N_f-N)$ sigma model. Following Witten \cite{W79,W93}, we describe its bosonic sector as a Higgs branch   of the following $U(1)$ linear gauged sigma model,
\begin{align}
	&S_{r=N} = \int d^2x \Bigg\{ \frac{1}{4e_0^2} F_{\alpha\beta}^2
	+ \frac{1}{e_0^2} |\partial_\alpha \sigma|^2+\sum_{P=1}^{N}\left(
	|\nabla_\alpha n^P|^2+|\sigma + m_P|^2 |n^P|^2\right)\nonumber
	\\
	&+\sum_{K=N+1}^{N_f}\left(|\tilde{\nabla}_\alpha \rho^{K}|^2 + |\sigma + m_N|^2 |\rho|^2\right)+
	\frac{e_0^2}{2}
	\left( \sum_{P=1}^{N}|n^P|^2 
	- \sum_{K=N+1}^{N_f}|\rho^K|^2 - 2\beta \right)^2
	\Bigg\}+S_{\text{def}}.
	\label{2D_action_r=N_bosonic}
\end{align}
Here, the 2D fields $n^{P}$ are associated with orientation moduli reflecting $SU(N)_{C+F}$ flux rotations, while the $\rho^K$ fields come from the size moduli. They have charges $+1$ and $-1$ under the gauge $U(1)$,
respectively. 
The complex scalar $\sigma$ is the superpartner of the gauge field. The limit $e_0\to \infty$ eliminates the dynamics of components of the gauge supermultiplet  and restores the condition    
\begin{align}
	\sum_{P=1}^{N}|n^P|^2 
	- \sum_{K=N+1}^{N_f}|\rho^K|^2 = 2\beta  \,.
	\label{couplings}
\end{align}
The 2D fields have twisted masses, $m_1,\ldots,m_N$ for $n^P$ and $m_{N+1},\ldots,m_{N_f}$ for $\rho^K$. 
These twisted mass parameters coincide with the mass parameters for the corresponding quark hypermultiplets in 4D, cf. Eq.~\eqref{N=2_superpotential_4D}.

The last term in the action \eqref{2D_action_r=N_bosonic} is a SUSY-breaking potential inherited from the single-trace deformation \eqref{def} in 4D. It has the form \cite{Shifman:2010kr,Bolokhov:2013bea}
\begin{align}
	S_{def} = \int d^2x \left\{4\pi |\mu \sigma|\right\}.
	\label{S_def}
\end{align}
In particular, this  2D deformation satisfies   the following requirement: the tensions of each of the  
$Z_N$  strings in 4D, given by 
\begin{align}
	T_{P}= 2\pi |\xi_{P}|, \quad P=1,\ldots,N;
	\label{tensions}
\end{align}
correspond to the values of the  2D scalar potential \eqref{S_def} at its minima \cite{Shifman:2010kr,Bolokhov:2013bea} .

This potential explicitly violates $\mathcal{N}=(2,2)$ supersymmetry.
Its effects will be discussed in Sec.~\ref{sec:kinkmasses} below.
For now, we will assume that $\mu$ is small, and focus on the leading $\mathcal{N}=(2,2)$ part of the 2D theory.

\begin{figure}[t!]
	\centering
	\includegraphics[width=0.8\textwidth]{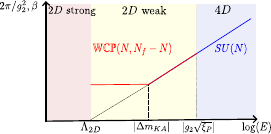} 
	\caption{RG flows of $\beta$ and $g_2$ couplings in the case of $r=N$. Equality $\beta=\frac{2 \pi}{g_2^2}$ is valid at the 
		$g_2\sqrt{\xi_P}$ scale. In the limit of $|\Delta m _{KP}| \gg \Lambda_{2D}$, the 2D theory develops a weak-coupling regime. }
	\label{fig:rg_flow1}
\end{figure}

The supersymmetric model 
\eqref{2D_action_r=N_bosonic} 
is asymptotically free.
Its beta function is one-loop exact; moreover, the first (and only) coefficient $b$
\begin{equation}
	b=N-(N_f-N)=2N-N_f
	\label{beta_func_2D_r=N}
\end{equation}
precisely coincides with that of the 4D theory.
The natural ultraviolet cut-off parameter of this effective theory   is of the order of the gauge boson masses $\sim g_2\sqrt{\xi_{P}}$. 
At that scale, we have equality of the 2D and 4D couplings \cite{SYrev}
\begin{equation}
	2\beta = \frac{4 \pi}{g_2^2}, \qquad \frac{8\pi^2}{g_2^2(E)}= b\log\frac{E}{\Lambda_{4D}}.
\end{equation}
This, in turn, implies equality of the scales of both theories (see Fig.~\ref{fig:rg_flow1} ) 
\begin{align}
	\Lambda_\text{4D} =\Lambda_\text{2D}.
	\label{Lambda_2D=4D}
\end{align}
In the limit $|\Delta m | \gg \Lambda_\text{2D}$, the 2D theory is in the weak-coupling regime.
This enables us to perform a quasiclassical analysis. 

Quasiclassically, for each $P_0=1,\ldots N$, there is an isolated vacuum 
\begin{align}
	n^{P}=\sqrt{2\beta}\,\delta^{P_0 P}, \quad \rho^{K}=0,  \quad \sigma=-m_{P_0}.
\end{align}
Since each of them is defined by a certain direction of $n^{P}$, these classical vacua correspond to $Z_N$ Abelian strings.  As we reduce $\Delta m$ and enter the strong-coupling region, the fields  $n^P$ and $\rho^K$ fluctuate over the whole target space \eqref{couplings}, and the string becomes truly non-Abelian. Still, the arguments based on the Witten index ensure that the number of vacua in the world sheet theory \eqref{2D_action_r=N_bosonic} remains equal to $N$.

\subsection{$r < N$ case}
\label{sec:r<N}

Much like in the $r = N$ case, the world sheet theory for $r < N$ is a sum of an $\mathcal{N}=(2,2)$ part and the deformation,
\begin{align}
	S_{2D}=S_{(2,2)}^{cl}+S_{def}.
	\label{2,2+def}
\end{align}
In order to determine $S_{(2,2)}^{\text{cl}}$ at the classical level, we follow the same strategy as used  in \cite{Shifman:2014lba} for the $r=N-1$ vacuum, but
generalize it to arbitrary $r$. 
Let us work under the same hierarchy of parameters as in Table~\ref{tab:scales_hierarchy_table_normal}.
In particular, for the moment, we assume that
\begin{align}
	|m_A|\gg|\sqrt{\xi_P}|\gg|\Delta m_{AB}|\gg \Lambda_{4D}, 
	\label{assump}
\end{align}
The gauge-group breaking has the following pattern
\begin{equation}
	U(N) \overset{|m_A|}{\xrightarrow{\hspace{1cm}}} U(N-r) \cross U(r) \overset{|\sqrt{\xi_P}|}{\xrightarrow{\hspace{1cm}}}U(N-r) \ldots 
	\label{gauge_breaking}
\end{equation}
Below the scale $\sim m_A$, all quarks with color index $k= (r+1),...,N$ acquire large masses $\sim m_A$ and decouple, see \eqref{phi}. As a result, the theory factorizes into \ntwo SQCD with gauge group $U(r)$ and $N_f$ quark flavors and pure \ntwo Yang-Mills theory with gauge group $U(N-r)$ without quarks. In the first one, the number of condensed quarks $r$ coincides with the rank of the gauge group, so it supports non-Abelian strings described by the world sheet theory $\mathbb{WCP}(r,N_f-r)$ with $r$ orientation moduli $n^P$,  together with $N_f-r$ size moduli $\rho^K$.

The beta function coefficient $b$ for this world-sheet theory is $b_{LE} = r- (N_f-r)=2r-N_f$; note that it coincides with the beta function of the $U(r)$ sector of the 4D gauge theory (in the corresponding range of energies $[m, \sqrt{\xi}]$).
The 2D fields have twisted masses, $m_1,\ldots,m_r$ for $n^P$ and $m_{r+1},\ldots,m_{N_f}$ for $\rho^K$. 
These twisted mass parameters coincide with the mass parameters for the corresponding quark hypermultiplets in 4D SQCD, cf. \eqref{2D_action_r=N_bosonic}.

Now we  relax the condition $|\Delta m_{AB}|\ll |m_A|$. Our goal is to find the appropriate \ntwot part of the 2D world sheet action for arbitrary $r$, see \eqref{2,2+def}.

Let us start from the  $r=N$-vacuum in 4D and set a number of the bare quark masses to zero, namely $m_{r+1}=\ldots =m_{N}=0$. 
At this special point, the  $r=N$ vacuum we started with merges with the  $r$-vacuum, which is evident from Eqs.~\eqref{phi} and \eqref{q}.
At this point, we can smoothly pass to the $r$-vacuum and then increase masses $m_{r+1},\ldots , m_{N}$ to their initial values. In this process, our world sheet theory is smoothly
deformed from the  $\mathbb{WCP}(N,N_f)$ model to the \ntwot world-sheet theory for the non-Abelian string in the $r$-vacuum we are looking for. Note, however,
that this process involves ``relabeling'' $(N-r)$ $n^A$ fields into $\rho^A$ fields with index $A=(r+1),\ldots,N$, which would change the $\beta$-function of the world sheet theory. In order to keep
the world-sheet $\beta$-function intact (it should coincide with the $\beta$-function of the 4D SQCD with $b=2N-N_f$),  we, following \cite{Shifman:2014lba}, add $2(N-r)$ extra fields with charge $+1$ and zero mass parameters, which we label $n^P$, $P=(r+1),\ldots,N$ and $z^A$, $A=1,\ldots,(N-r)$.

Combining these features with $\mathcal{N}=(2,2)$ symmetry, we finally arrive at the following 2D action: 
\begin{align}
	S^{\text{cl}}_{(2,2)} =& \int d^2x \Bigg\{ \frac{1}{4e_0^2} F_{\alpha\beta}^2
	+ \frac{1}{e_0^2} |\partial_\alpha \sigma|^2 \nonumber \\
	&+\sum_{P=1}^{r}\left(
	|\nabla_\alpha n^P|^2+|\sigma + m_P|^2 |n^P|^2\right) 
	+\sum_{K=r+1}^{N_f}\left(|\tilde{\nabla}_\alpha \rho^{K}|^2+ |\sigma + m_K|^2 |\rho^K|^2\right) \nonumber \\
	&+\sum_{A=1}^{N-r}\left( |\nabla_\alpha n^{r+A}|^2+|\sigma|^2 |n^{r+A}|^2
	+ |\nabla_\alpha z^{A}|^2+|\sigma|^2 |z^A|^2\right)\nonumber 
	\\
	&+ \frac{e_0^2}{2}
	\left( \sum_{P=1}^{r}|n^P|^2 +\sum_{A=1}^{N-r}\left( |z^A|^2 +|n^{r+A}|^2\right)- \sum_{K=r+1}^{N_f}|\rho^K|^2 - 2\beta \right)^2
	\Bigg\}.
	\label{2D_model_unwinding}
\end{align}
%
The coefficient $b$ of the $\beta$-function of this 2D model is equal to the number of fields with charge $+1$ minus the number of fields with charge $-1$, $b=r  + 2(N-r) -(N_f-r)=2N-N_f$.

\begin{figure}[t!]
	\centering
	\includegraphics[width=0.9\textwidth]{
		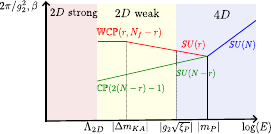} 
	\caption{  RG flows of $\beta$ and $g_2$ couplings in the case of $r<N$ and $N_f>2r$. The $SU(r)$ part of the IR theory interacts with $N_f$ flavors, while quarks from the $SU(N-r)$ sector are decoupled. Thus, $b_{SU(r)}=2r-N_f$ and $b_{SU(N-r)}=2(N-r)$, which coincide with the $b$ coefficients  of $\mathbb{WCP}(r,N_f-r)$ and $\mathbb{CP}(2(N-r)-1)$.} 
	\label{fig:rg_flow2}
\end{figure}

It is instructive to verify our result and reproduce our conclusion in the limit \eqref{assump}. Consider $r$ classical vacua of the model \eqref{2D_model_unwinding}:
\begin{align}
	\rho^{K}=0, \quad z^A=0, \quad n^{P}=\sqrt{2\beta}\,\delta^{P_0 P},  \quad \sigma=-m_{P_0},
	\label{classical_n}
\end{align}
where $P_0=1,\ldots,r$. 
In these vacua, the fields $z^A$ and $ n^{r+1},\ldots,n^{N}$ are heavy and decouple, leaving us with the $\mathbb{WCP}(r,N_f-r)$, as expected (see also Fig.\ref{fig:rg_flow2}).

Another set of (classical) vacua corresponds to
\begin{align}
	\sum_{A=1}^{N-r}\left( |z^A|^2 +|n^{r+A}|\right)=2\beta, \quad \sigma=0,
\end{align}
All the other fields become heavy and decouple. Thus, at the classical level, we obtain a massless $\mathbb{CP}(2(N-r)-1)$  model.  This part of the world sheet theory is strongly modified at the quantum level by the presence of the gaugino condensate in 4D SQCD. We will see below that  the extra vacua associated with this part describe electric strings.

We perform a detailed analysis of the quantum effects in the world sheet theory in   Sec.~\ref{sec:Quantum_world_sheet}.
The role of the deformation term $S_{\text{def}} $ in \eqref{2,2+def}, see Eq.~\eqref{S_def}, is discussed in Sec.~\ref{sec:kinkmasses}.

\section{Confinement of quarks and monopoles}
\label{sec:confiniment}

As was mentioned above, in the hybrid $r$-vacua of  $\mathcal{N}=2$ SQCD   non-Abelian strings are responsible for the confinement of  monopoles and quarks belonging to orthogonal sectors of the gauge group.
In this section we draw a qualitative picture of confinement and hadronic states in this theory at the quasiclassical level.
Quantum aspects are described in the later sections of this paper.

\subsection{Basic features of  monopoles}

Let us start with the Coulomb branch of the $\mathcal{N}=2$ gauge theory with the deformation \eqref{def} switched off, $\mu = 0$.
In this setup, the squarks do not form condensates \eqref{q}, and the vacua are determined solely by the adjoint VEV.
The latter satisfies
\begin{equation}
	[a_a T^a,\, a_c T^c] = 0 \,,
\end{equation}
which implies that the matrix \( a_a T^a \) lies in the Cartan subalgebra of the UV gauge group, $ a_a T^a =\vec{h}\cdot\vec{H}.$ 
As a result, the gauge group is spontaneously broken as $SU(N) \;\longrightarrow\; U(1)^{N-1}$,
leading to the appearance of \( N - 1 \) elementary ’t~Hooft–Polyakov monopoles \cite{SW,SW2}.

Each of the monopoles is associated with a simple root $\vec{\gamma}_j$ of the algebra  $\mathfrak{su} (N)$. Using weights $\vec{w}_{j}$ in the fundamental representation, one can write the roots as $\vec{\gamma}_j =\vec{w}_{j}-\vec{w}_{j+1}$, where $j=1,\ldots,N-1$ and $\vec{H}q^j = \vec{w}_jq^j$, so we label these monopoles (and sometimes their masses) as $M_{12},M_{23},\ldots,M_{N-1 N}$.  Their charges and masses in the classical limit  are determined by the following relations
\begin{align}
	\vec{\mathtt{Q}}^\text{mag}_{j,j+1}= \vec{\gamma_j}, \quad M_{j,j+1}^{\mathrm{Coulomb}}= \frac{4\sqrt{2}\pi}{g_2^2} \left(\vec{\mathtt{Q}}^\text{mag}_{j,j+1} \cdot \vec{h}\right). \label{tpmasses}
\end{align}
In fact, for every root 
$\vec{\alpha} = \sum\limits_{j} c_{j}\vec{\gamma}_j$ there is a composite solution whose quantum numbers form the corresponding linear combination of the quantum numbers of the elementary monopoles.

Now, let us switch the $\mu \neq 0$ deformation back on, but keep it small, so that $|\Delta m_{KA}|\gg \sqrt{\xi_P} $ (here, $\Delta m_{KA} = m_K - m_A$ are the quark mass differences).  The Coulomb branch gets lifted.  Below the quark mass scale $m_A$ the gauge group is broken as shown in \eqref{gauge_breaking} and 
$r$ squarks develop VEVs in the  $U(r)$ SQCD. This leads to the formation of non-Abelian strings in the $U(r)$ sector, which confine $r$ elementary monopoles with masses given by
\beq
M_{j,j+1}^{\mathrm{Coulomb}}= \frac{4\pi |m_{j+1} -m_j|}{g_2^2}, \qquad j=1,...,(r-1), \quad 
M_{r,r+1}^{\mathrm{Coulomb}}= \frac{4\pi |m_r|}{g_2^2}
\label{mon_mass_cl}
\eeq 
on the Coulomb branch; see \cite{SYrev}.
In the remaining pure Yang-Mills $U(N-r)$ theory,   the elementary massless monopoles $M_{r+1,r+2} \ldots M_{N-1,N}$  condense  by the Seiberg-Witten mechanism \cite{SW,SW2}.

To illustrate this more explicitly, and at the same time to show that the electric strings resulting from monopole condensation confine quarks, we consider the following hierarchy of parameters 
\begin{align}
	|m_P|\gg|\Delta m_{KA}|\gg \Lambda_{4D}\gg|\sqrt{\mu m_P }|\gg |\sqrt{\mu \Lambda_{4D}}|,
\end{align}
see Table~\ref{tab:scales_hierarchy_table_2} for more details.
The $U(1)^N$ gauge group breaks further by the VEVs of squarks and monopoles down to a single $U(1)_{\text{unbr}}$. 
As we already mentioned, the degrees of freedom that become light are still mutually local, so there is no issue with the Lagrangian description (see e.g. \cite{Shifman:2013zsa}).

For illustration, let us consider two cases.

\begin{table}[t!]
	\centering
	\begin{tabular}{|c|c|c|}
		\hline
		\textbf{Energy scale} & \textbf{Gauge symmetry} & \textbf{Flavor-related symmetry} \\
		\hline
		$m \ll E$  &  $U(N)$  & $SU(N_f)_\text{flavor}$ \\
		\hline \rowcolor{gray!15} 
		$E \sim m$ & \multicolumn{2}{|c|}{Adjoint scalar condensation (classical)} \\
		\hline
		$\Delta m \ll E \ll m$  &  $U(r) \times  U(N-r)$  & $SU(N_f)_\text{flavor}$ \\
		\hline \rowcolor{gray!15} 
		$E \sim \Delta m$ & \multicolumn{2}{|c|}{Flavor symmetry breaking}  \\
		\hline
		$\Lambda \ll E \ll \Delta m$  &  $U(1)^r \times  U(N-r)$  & $U(1)^{N_f-1}$  \\
		\hline \rowcolor{gray!15} 
		$E \sim \Lambda$ & \multicolumn{2}{|c|}{ Adjoint scalar condensation (non-pert) }  \\
		\hline
		$\sqrt{\xi} \sim \sqrt{\mu m} \ll E \ll \Lambda$  & $U(1)^{N}$  & $U(1)^{N_f-1}$ \\
		\hline \rowcolor{gray!15} 
		$E \sim \sqrt{\xi}$ & \multicolumn{2}{|c|}{Squark condensation} \\
		\hline
		$\sqrt{\mu \Lambda} \ll E \ll \sqrt{\xi}$  &  $U(1)^{N-r}$  & $U(1)^{N_f-1}$ \\ 
		\hline \rowcolor{gray!15} 
		$E \sim \sqrt{\mu \Lambda}$ & \multicolumn{2}{|c|}{ Monopole condensation }  \\
		\hline
		$E \ll \sqrt{\mu \Lambda}$ & $U(1)_\text{unbr}$ & $U(1)^{N_f-1}$ \\
		\hline
	\end{tabular}
	\caption{Scale hierarchy for generic $r$ }
	\label{tab:scales_hierarchy_table_2}
\end{table}

\subsection{Gauge group $U(3)$, the $r=1$ vacuum}
\label{sec:example_U3_r1}

This theory is a minimal example of SQCD with $N_f\ge 3$ quark flavors in a   hybrid vacuum with a single condensed monopole and a single condensed squark. 
In the deep IR, the squark $q_{11}$ and the monopole $M_{23}$ develop VEVs, breaking the gauge group to just $U(1)_\text{unbr}$. In this subsection we review this example considered in \cite{Shifman:2013zsa} using a different basis for the $U(3)$ generators; see also \cite{SYrvacua}, where confinement of monopoles in SQCD with gauge group $U(4)$ in the $r=3$ vacuum was considered.

\subsubsection{$U(1)^3 \to U(1)_\text{unbr}$ effective theory}

For a moment, let us, however, discuss the theory at slightly higher energies.
At the energy range between $\Lambda_{4D}$ and $\xi_P$, the theory contains three \( U(1) \) gauge fields.
One can choose a basis in the Cartan subalgebra of $\mathfrak{u} (3)$ as
\begin{align}
	H_1=\text{diag}(a,0,0), \quad H_{23}=\text{diag}(0,b,-b), \quad H_{\text{unbr}}=\text{diag}(0,b,b),
	\label{example_r=1_cartan_basis_1}
\end{align}
These generators are orthogonal and normalized (with the convention Eq.~\eqref{algebra_normalization}) provided that $a^2=1/2$, $b^2=1/4$. 
However, at intermediate steps of the calculation below, we keep the parameters $a$, $b$ general.

The monopoles are determined only by the non-Abelian part of the UV gauge group, which means that their magnetic charges are determined only by the generators from the Cartan subalgebra of the $\mathfrak{su} (3)$ part, and not the full $\mathfrak{u} (3)$.
The relevant subalgebra is the span of $H_{12}$ and $H_{23}$, where 
\begin{align}
	H_{12}=\frac{1}{c}\left( H_1+H_{\text{unbr}}-\frac{\tr(H_1+H_{\text{unbr}})}{3}\right)= \frac{2(a-b)}{3ac}H_1+\frac{b-a}{3bc}H_{\text{unbr}},
\end{align}
with a normalization condition $\frac{2(a-b)^2}{3c^2}=1/2$. 

In order to determine the electric and magnetic charges of the monopoles and quarks, we need to write down the weights and roots.
The weights of the  $\mathfrak{su} (3)$ part are 
\begin{align}
	\Vec{w_1}=\left(\frac{a}{c}-\frac{2b+a}{3c},0\right), \quad \Vec{w_2}=\left(\frac{b}{c}-\frac{2b+a}{3c},b\right), \quad  \Vec{w_3}=\left(\frac{b}{c}-\frac{2b+a}{3c},-b\right),
\end{align}
and using them we can find the simple roots 
\begin{align}
	\vec{ \gamma_1}=\vec{w_1}-\vec{w_2}=\left(\frac{a-b}{c},-b\right), \quad \vec{\gamma_2}=\vec{w_2}-\vec{w_3}=\left(0,2b\right).
	\label{example_r=1_roots}
\end{align}
We also introduce our notation for the set (a ``vector'') of electric and magnetic charges with respect to the \( U(1)^3 \) gauge fields,
\begin{align}
	\vec{n}_{q} =\left(\mathtt{Q}^{\text{el}}_{1},\mathtt{Q}^{\text{el}}_{23},\mathtt{Q}^{\text{el}}_{\text{unbr}} \right); \quad
	\vec{m}_{M}=(\mathtt{Q}^{\text{mag}}_{1},\mathtt{Q}^{\text{mag}}_{23},\mathtt{Q}^{\text{mag}}_{\text{unbr}}) \,.
\end{align}
The subscripts refer to the three respective Cartan generators from Eq.~\eqref{example_r=1_cartan_basis_1}.

As mentioned above, in the deep IR, the squark $q_{11}$ and the monopole $M_{23}$ develop VEVs.
The condensed states carry the following electric and magnetic charges, respectively,
\begin{align}
	&\vec{n}_{q^{11}}=(a,0,0); \quad \vec{m}_{M_{23}}=\left(0,2b,0\right); 
	\label{example_r=1_charges}
\end{align}
Here, the quark's electric charge is determined by the corresponding weight (see Eq.~\eqref{example_r=1_cartan_basis_1} and Appendix~\ref{sec:appendix_charges}), while the monopole charge follows from Eqs.~\eqref{tpmasses} and \eqref{example_r=1_roots}.
As one can see, each of these states interacts only with a single combination of the $U(1)$ gauge fields
\begin{align}
	q^{11}: \ a A^1_{\mu}, \quad M_{23}: \ 2b A^{D 23}_{\mu}.
\end{align}
Here, $A^{D 23}_{\mu}$ is the dual gauge field for the $U(1)$ gauge group which is aligned with the generator $H_{23}$ in Eq.~\eqref{example_r=1_cartan_basis_1}.

\subsubsection{Electric and magnetic strings}

In the presence of the scalar condensates, one can construct Abelian strings.
Here, they come in two kinds, corresponding to the winding of the squark or the monopole:
\begin{align}
	q^{11}\sim e^{i \alpha}\sqrt{\frac{\xi_1}{2}} \quad \text{or} \quad M_{23} \sim  e^{i \alpha}\sqrt{\frac{\xi_2}{2}}; \quad r_\perp \to \infty, 
	\label{example_r=1_winding_Abelian}
\end{align}
Here, $\alpha$ and $r_\perp$ are the polar coordinates in the plane $(x_1,x_2)$ orthogonal to the string (the string is taken to be straight and infinite).
Correspondingly, the winding \eqref{example_r=1_winding_Abelian} together with the set of charges \eqref{example_r=1_charges} determines the asymptotics of the gauge fields,
\begin{align}
	a A^{1}_k  \sim \partial_{k}\alpha\quad \text{or} \quad 2b A^{D 23}_k  \sim \partial_{k}\alpha; \quad |x|\to \infty,
	\label{example_r=1_string_gauge_asympt}
\end{align}
while other components are zero. These asymptotics ensure that the covariant derivatives of the condensed fields $\nabla_i q^{11}$ and $\nabla_i M_{23}$, $i=1,2$, fall off faster than $1/r_\perp$ at large $r_\perp$, so the corresponding string tensions are finite.

The asymptotics \eqref{example_r=1_string_gauge_asympt} allow us to compute the gauge field fluxes carried by each of the strings,
\begin{align}
	\vec{S}_{m} =\frac{1}{4\pi}\oint dx^k \left(A^{1}_k,A^{23}_k,A_k^{\text{unbr}}\right); \quad \vec{S}_{e} =-\frac{1}{4\pi}\oint dx^k \left(A^{D1}_k,A^{D23}_k,A_k^{D\text{unbr}}\right).
	\label{def_string_charge}
\end{align}
Here, gauge fluxes  define the string charges equal to the probe charges of monopoles or quarks, which  can be attached to the endpoints of the string \cite{Shifman:2013zsa}.  Note that this probe charge does
not necessarily exist in the theory under consideration.
Substituting Eq.~\eqref{example_r=1_string_gauge_asympt} here, we obtain
\begin{align}
	\Vec{S}^{1}_{m}=\left(\frac{1}{2a},0,0\right); \quad \Vec{S}^{23}_{e}=(0,- \frac{1}{4b},0)  \,.
\end{align}

\subsubsection{Mesonic states}
\label{sec:mesonic_1}

\begin{figure}
	\includegraphics[width=1\textwidth]{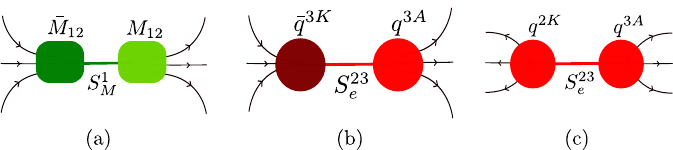} 
	\caption{Dumbbell configurations formed by squarks and  monopoles. Circles represent quarks, and squares represent monopoles. Light and dark shapes denote particles and antiparticles, respectively. Green and red lines correspond to electric and magnetic strings. Black lines depict the field lines of $U(1)_{\text{unbr}}$.  }
	\label{fig:mesons_u3_r1}
\end{figure}

Now, let us turn our attention to the fields that stay massive and do not form condensates, namely $q^{2A},q^{3A},M_{12},M_{13}$.
As we will see shortly, these states have electric and magnetic charges with respect to the two broken $U(1)$'s; as a result, they become confined by the flux tubes (strings) just discussed.

The charges of the heavy quark (monopole) can be decomposed into the screened component proportional to $\vec{n}_{q^{11}}$ ($ \vec{m}_{M_{23}}$), the unbroken component proportional to $\vec{n}^{\text{unbr}}$ ($\vec{m}^{\text{unbr}}$), and the confined component  equal to $\pm \vec{S}_{e}^{23}$($\pm\vec{S}_{m}^{1})$. 
Namely, the quarks $q^{2A},q^{3A}$ carry the charges (see also Appendix~\ref{sec:appendix_charges})
\begin{align}
	&\vec{n}_{q^{2A}} =(0,b,b)=-  \vec{S}^{23}_{e}+\frac{1}{2}\vec{n}^{\text{unbr}}, \nonumber \\
	& \vec{n}_{q^{3A}} =(0,-b,b)= \vec{S}^{23}_{e}+\frac{1}{2}\vec{n}^{\text{unbr}}.
	\label{example_r=1_charges_heavy_quarks}
\end{align}
Similarly, for the monopoles $M_{12},M_{13}$ we obtain 
\begin{align}
	&\vec{m}_{M_{12}}= \frac{a-b}{c}\vec{m}^{12}-b\vec{m}^{23}=\vec{S}^{1}_m-\frac{1}{2}\vec{m}_{M_{23}}-\frac{1}{2}\vec{m}^{\text{unbr}}  \,, \nonumber \\ 
	& \vec{m}_{M_{13}}= \frac{a-b}{c}\vec{m}^{12}+b\vec{m}^{23}= \vec{S}^{1}_m+\frac{1}{2}\vec{m}_{M_{23}}-\frac{1}{2}\vec{m}^{\text{unbr}},
	\label{example_r=1_charges_heavy_monopoles}
\end{align}
where in \eqref{example_r=1_charges_heavy_quarks} and \eqref{example_r=1_charges_heavy_monopoles} we use the numerical values for $a,b$ and $c$.

This simple example illustrates the emergence of ``endpoint'' states, i.e. the states on which the strings can end. 
Clearly, the quarks $q^{2A}$ and $q^{3A}$ have a charge component that has to be confined ($\vec{S}^{23}_{e}$ in Eq.~\eqref{example_r=1_charges_heavy_quarks}), and so they must be attached to opposite ends of an electric string, that is, the string carrying an electric field flux.
Similarly, as follows from Eq.~\eqref{example_r=1_charges_heavy_monopoles}, a magnetic string can have $M_{13}$ and $\bar{M}_{12}$ (or $\bar{M}_{13}$ and $M_{12}$, where the bar denotes an antiparticle) at its endpoints\footnote{Note that quarks  $q^{1A}$ and  the monopole $M_{23}$  are just screened and can be ignored in the present analysis.}.
Of course, any three-dimensional vector of the quark or monopole  charges can
always be written as a linear combination of three orthogonal vectors. What is nontrivial in 
\eqref{example_r=1_charges_heavy_quarks} and \eqref{example_r=1_charges_heavy_monopoles}  is the coefficient in front of the string charge: it should be an integer to ensure confinement.
Consequently, the spectrum of the bulk states includes color-neutral composite states (mesons) depicted in Fig. \ref{fig:mesons_u3_r1}.

Each dumbbell formed by a particle-antiparticle pair represents a dipole configuration emitting fluxes of  the unbroken $U(1)$ \cite{Shifman:2013zsa,SYrvacua}. 
In turn, the dumbbell composed of $q^{2A}$ and $q^{3B}$ appears as a charged stringy meson. 

Note that since the charge with respect to the generator $H_{23}$ is completely screened and the composite monopole $M_{13}$ has the same mass as $M_{12}$, these two constituents are physically indistinguishable in the IR effective theory.

Finally, let us note the representations of these mesons with respect to the global symmetries.
If we push the scale of the quark mass differences $\Delta m$ below $\Lambda$ and $\sqrt{\xi}$, we effectively restore the $SU(N_f-r)_\text{fl}$ part of the flavor symmetry as well as the $SU(r)_\text{C+F}$ color-flavor-locked symmetry.
In our case, the color-flavor-locked part is trivial, while the flavor group, say for $N_f=3$, is $SU(2)_\text{fl}$.
Then, the quark-quark mesons from Fig.~\ref{fig:mesons_u3_r1} form multiplets belonging to 
$\mathbf{2} \otimes \mathbf{2} \cong \mathbf{1} \oplus \mathbf{3}$ 
with respect to the flavor symmetry.
Similarly, the quark-antiquark mesons from Fig.~\ref{fig:mesons_u3_r1} transform under
$\mathbf{2}\otimes \overline{\mathbf{2}} \cong \mathbf{1}\oplus \mathbf{3}$.

In this  example we see the emergence of pairs of particles, each confined by one elementary string.
The resulting configurations resemble a simple meson.
In Appendix \ref{U_5_r_2} we consider  a more complicated scenario, where several particles are confined by a closed   non-Abelian string forming a ``necklace'' configuration (as always happens in $r=N$ vacua to monopoles, see \cite{SYrev}), or where more than two particles become confined by an open string.

\section{Quantum deformation of the world-sheet theory }
\label{sec:Quantum_world_sheet}

In the previous section, we used the quasiclassical approximation assuming  large $\Delta m \gg \Lambda_{4D}$.
In this limit, the orientations $n^P$ are fixed and the non-Abelian string reduces to $Z_N$ strings associated with classical vacua of the world-sheet model, cf. \eqref{classical_n}. In this regime
we discussed the classical picture of monopole and quark confinement in the hybrid $r$-vacuum. 

Now we will switch on quantum effects using an exact description, which 
remains valid even beyond all quasiclassical approximations employed earlier. In this regime, quantum effects are enhanced, rendering the quasiclassical analysis inapplicable. To extend our description of confinement into this domain, we therefore return to the two-dimensional effective theory defined on the world-sheet of the non-Abelian string
(see Sec.~\ref{sec:r<N}).

The action \( S_{(2,2)}^{\text{cl}}\) in \eqref{2D_model_unwinding} is classical. At the quantum level,  a non-vanishing gluino condensate \eqref{glu_cond} in $4D$ SQCD affects the target space of the world-sheet theory, generating non-perturbative effects that lie ``outside'' of the original classical model. 
We avoid the question of directly modifying the target space and instead use another approach based on the exact effective superpotential of the Veneziano-Yankielowicz type   \cite{Veneziano:1982ah}.

\subsection{Exact   effective superpotential in the $r=N$ vacuum }
\label{sec:sup_r=Nl}

An  exact twisted superpotential  is known  for
the \cpn model \cite{AdDVecSal,Cecotti:1992rm,W93,Dorey}.
This superpotential was later generalized to the case of the $\mathbb{WCP}$ models in 
\cite{HaHo}.
Integrating out the fields $n^P$ and $\rho^K$  in \eqref{2D_action_r=N_bosonic} we obtain
the following
exact twisted superpotential:
\beqn
&&{\cal W}_{\rm WCP}(\Sigma)= 
\frac1{4\pi}\left\{\sum_{P=1}^N\,
\left(\Sigma+{m}_P\right)
\,\ln{\frac{\Sigma+{m}_P}{\Lambda_{4D}}}
\right.
\nonumber\\[3mm]
&& 
\left.
-\sum_{K=N+1}^{N_f}\,
\left(\Sigma+{m}_K\right)
\,\ln{\frac{\Sigma+{m}_K}{\Lambda_{4D}}} 
- (2N-N_f) \,\Sigma \right\}\, ,
\label{CPsup}
\eeqn
where we  have introduced the twisted chiral superfield \( \Sigma \); its lowest component is the complex scalar $\sigma$. We also  used the condition $\Lambda_{2D}=\Lambda_{4D}$, see \eqref{Lambda_2D=4D}.
Minimizing this superpotential with 
respect to $\sigma$, we get the equation for the VEVs of $\sigma$ (the so-called twisted chiral ring equation),
\beq
\prod_{P=1}^N(\sigma+{m}_P)
=\Lambda_{4D}^{(2N-N_f)}\,\prod_{K=N+1}^{N_f}(\sigma+{m}_K)\,.
\label{sigmaeq}
\eeq

The  masses of the BPS kinks interpolating between the
vacua $\sigma_{P}$ and $\sigma_{P'}$ are given  by the appropriate 
differences of the superpotential (\ref{CPsup}) calculated at distinct roots of \eqref{sigmaeq} \cite{HaHo,Dorey,DoHoTo},
\beq
M^{\rm BPS}_{PP'} =
2\left|{\cal W}_{\rm WCP}(\sigma_{P'})-{\cal W}_{\rm WCP}(\sigma_{P})\right|\,,\qquad P,P'=1,..., \,N\,.
\label{BPSmass}
\eeq
Due to the presence of branches  in the logarithmic functions in (\ref{CPsup}) each kink comes together with a 
tower of dyonic kinks carrying global U(1) charges (for more details
see e.g. \cite{Bolokhov:2012dv}).
The  masses obtained from (\ref{BPSmass}) were shown  
to coincide with those of the monopoles and dyons in the bulk theory. The latter are 
given by the period integrals of the Seiberg--Witten curve (\ref{swcurve}). 

As was mentioned in the Introduction,
this coincidence was observed in \cite{Dorey,DoHoTo} and   explained later 
in \cite{SYmon,HT2} using the picture of confined bulk monopoles which are seen as kinks in the world 
sheet theory. A crucial point is that both monopoles and kinks are BPS-saturated states\,\footnote{Confined
	monopoles, being junctions of two distinct 1/2-BPS strings, are 1/4-BPS states in the bulk theory 
	\cite{SYmon}.},
and their masses cannot depend on the non-holomorphic parameter $\xi$ \cite{SYmon,HT2}. This means that,
although confined monopoles look physically very different from unconfined monopoles on the Coulomb branch
of the 4D SQCD (in the particular singular point which becomes an isolated vacuum at nonzero $\xi$),
their masses are the same. Moreover, they coincide with the masses of kinks in the world-sheet 
theory.

Note that the roots of the vacuum  equation (\ref{sigmaeq}) coincide with the double roots of the 
Seiberg--Witten curve (\ref{swcurve}) of 
the bulk theory \cite{Dorey,DoHoTo},
\beq
\sigma_P=\sqrt{2}\,e_P\, .
\label{equalroots}
\eeq
This is the key technical reason that leads to the coincidence of the BPS spectra. In 
\cite{Shifman:2014lba} it was shown that the same correspondence is also valid for the $r=N-1$ vacuum. We will see below that this relation can be generalized to all $r$ vacua.

\subsection{Exact effective  superpotential  in the $r$ vacuum}
\label{sec:Exact_twisted_superpotencial}

Now let us consider the effective twisted superpotential for the world-sheet theory \eqref{2D_model_unwinding} on  the non-Abelian string in the $r$ vacuum.
Collecting the contributions from the \( r \) orientational fields \( n^P \), the \(2( N-r) \) massless fields \( z^A \), \( n^{r+1},\ldots,n^N \), and the \( N_f - r \) size moduli fields, we obtain
\begin{align}
	\mathcal{W}&=\frac{1}{4\pi}\Bigg[\sum_{P=1}^{r}(\Sigma+m_P)\log\left( \frac{\Sigma+m_P}{e\Lambda_{4D}}\right)+2(N-r)\Sigma\log\left( \frac{\Sigma}{e\Lambda_{4D}}\right) \nonumber\\
	&-\sum_{K=r+1}^{N_f}(\Sigma+m_K)\log\left( \frac{\Sigma+m_K}{e\Lambda_{4D}}\right)\Bigg],
	\label{WVY}
\end{align}
Using the definition \eqref{phi}, we can rewrite the  superpotential \eqref{WVY} in terms of $\Phi_{\text{cl}}$  as \cite{Shifman:2014lba}
\begin{align}
	\mathcal{W}_{VY}(\Sigma)=\frac{1}{4 \pi}\left\{2 \operatorname{Tr}\left[\left(\Sigma-\sqrt{2} \Phi_{\mathrm{cl}}\right) \log  \frac{\Sigma-\sqrt{2} \Phi_{\mathrm{cl}}}{e \Lambda_{4D}}\right]\right.
	-  \left.\sum_{A=1}^{N_f}\left(\Sigma+m_A\right) \log \frac{\Sigma+m_A}{e \Lambda_{4D}}\right\}.
\end{align}
Thanks to the $\mathcal{N}=(2,2)$ supersymmetry, there are no higher-loop corrections to this formula.
According to  the approach of Gaiotto, Gukov, and
Seiberg \cite{Gaiotto:2013sma}, nonperturbative corrections to this superpotential can be taken into account by writing 
\begin{align}
	\mathcal{W}_{eff}(\Sigma)=\frac{1}{4 \pi}\left\{2 \left\langle \operatorname{Tr}\left[\left(\Sigma-\sqrt{2} \Phi \right) \log  \frac{\Sigma-\sqrt{2} \Phi}{e \Lambda_{4D}}\right] \right\rangle\right.
	-  \left.\sum_{A=1}^{N_f}\left(\Sigma+m_A\right) \log \frac{\Sigma+m_A}{e \Lambda_{4D}}\right\},
\end{align}
where the expectation value is taken in the 4D SQCD. Considering the second derivative, we obtain
\begin{align}
	\partial^2_{\Sigma}\mathcal{W}_{eff}=\frac{1}{4\pi}\left\{2 \left\langle \Tr\frac{1}{\Sigma-\sqrt{2}\Phi} \right\rangle-\sum_{A=1}^{N_f}\frac{1}{\Sigma+m_A}
	\right\}. 
	\label{d^2W}
\end{align}
The exact solution for the resolvent in \eqref{d^2W} was found by Cachazo, Seiberg, and Witten~\cite{Cachazo:2003yc}
\begin{align}
	\left\langle \Tr\frac{1}{x-\Phi} \right\rangle=\frac{P'(x)}{y(x)}-\frac{P(x)Q'(x)}{2y(x)Q(x)}+\frac{Q'(x)}{2Q(x)}, \label{resolvent}
\end{align}
where  $P(x)=\prod_{k=1}^{N}\left(x-\phi_{k}\right)  ,  Q(x) = \prod_{P=1}^{N_f}\left(x+\frac{m_P}{\sqrt{2}}\right)$, and $y(x)$ denotes the value on the principal sheet of the Riemann surface \eqref{swcurve}. To obtain the non-perturbatively exact superpotential, we substitute Eq.~\eqref{resolvent} into Eq.~\eqref{d^2W} and integrate with respect to $\Sigma$. Integrating Eq.~\eqref{d^2W} once and choosing suitable boundary conditions, we obtain
\begin{align}
	4\pi\partial_{\Sigma}\mathcal{W}_{eff}
	=
	2 \int^{\Sigma} dx \left\langle \left\{\Tr\frac{1}{x-\sqrt{2}\Phi}\right\} \right\rangle
	-\sum_{A=1}^{N_f}\log\left(\Sigma+m_A \right)
	-\log\Lambda_{4D}^{2N-N_f}.
	\label{dW_1}
\end{align}
Thus, 4D quantum effects enter the two-dimensional sigma model through the gauge-invariant parameters $\phi_k$. 
For the hybrid $r$-vacuum considered here, the generic solution for the resolvent \cite{Cachazo:2003yc} can be written \cite{Shifman:2014lba} as follows  (see  Appendix~\ref{resolvent_in_r} for the detailed derivation):
\begin{align}
	\left\langle\Tr\frac{1}{\Sigma-\sqrt{2}\Phi}\right\rangle  = &\frac{1}{2} \sum_{A=1}^{N_f}\frac{1}{\Sigma+m_A}+\frac{1}{2}\frac{2N-N_f}{\sqrt{\Sigma^2- \frac{4 S}{\mu}}}\nonumber\\ -&\frac{1}{2}\sum^r_{A=1} \frac{\sqrt{m_A^2-\frac{4S}{\mu}}}{\sqrt{\Sigma^2-\frac{4S}{\mu}}(\Sigma+m_A)}+\frac{1}{2}\sum^{N_f}_{A=r+1} \frac{\sqrt{m_A^2-\frac{4S}{\mu}}}{\sqrt{\Sigma^2-\frac{4S}{\mu}}(\Sigma+m_A)}.
	\label{exact_resolvent}
\end{align}
In this representation, the entire effect of the non-perturbative deformation from the bulk is encoded in the single parameter $S/\mu$.  While this parameter is defined for the 4D theory, in Sec.~\ref{sec:small_and_S} we show that it can be  found  entirely within the 2D theory.

One particularly simple case is the $r=N$ vacuum. In 4D SQCD, the gaugino condensate  vanishes $S=0$, and
it is straightforward to see that 
the exact superpotential reduces to the one in \eqref{CPsup} and the world-sheet theory remains undeformed by the bulk instantons.

Using Eq.~\eqref{exact_resolvent}, we can straightforwardly integrate Eq.~\eqref{dW_1}:
\begin{align}
	\partial_{\Sigma} \mathcal{W}_{eff}=&\frac{1}{4\pi}\Bigg\{(2N-N_f)\log\left( \frac{t}{\Lambda_{4D}}\right)-\sum_{A=1}^{r}\left[\log\left(\frac{t_A}{\Lambda_{4D}} \right)-\log\left( \frac{\Sigma+m_A}{\Lambda_{4D}} \right) \right] \nonumber \\ 
	+&\sum_{A=r+1}^{N_f}\left[\log\left(\frac{t_A}{\Lambda_{4D}} \right)-\log\left( \frac{\Sigma+m_A}{\Lambda_{4D}} \right) \right]\Bigg \}, 
	\label{dW}
\end{align}
where the quantities $t$ and $t_A$ are defined as follows
\begin{equation}
	\begin{aligned}
		2t_A &=\sqrt{\sigma^2-\frac{4 S}{\mu}}+\frac{\sigma+\frac{4 S}{\mu m_A}}{\sqrt{1-\frac{4 S}{\mu m_A^2}}}, \\
		2t &= \sqrt{\sigma^2-\frac{4 S}{\mu}}+\sigma.
	\end{aligned}
	\label{t_t_A}
\end{equation}
This equation represents the exact effective twisted superpotential for the \ntwot supersymmetric part of the world-sheet theory  for the non-Abelian string in the $r$ vacuum. It takes into account the quantum
deformation produced by the bulk instantons. The latter
generate a gaugino condensate that results in the emergence
of a square-root cut in the $\sigma$  plane, see \eqref{t_t_A}. The emergence of this cut is a response of the 2D  world-sheet theory to the cut in the SW curve
present in the 4D theory in the $r$  vacuum, see \eqref{SW_curve_rep-01}. Also, much in the same way as for the superpotential \eqref{CPsup} in the $r=N$ vacuum, logarithmic functions in \eqref{dW} have branch points at $\sigma = -m_A$, leading to towers of dyonic kinks. 

Note that the K\"ahler potential is subject to perturbative and non-perturbative corrections, and therefore the full \ntwot world-sheet action cannot be described by an exact formula.

\subsection{Chiral ring equation}
\label{sec:Chiral-ring_equation}

Our goal now is to determine the vacua of the 2D theory, which are given by the critical points of the superpotential:
\begin{align}
	\partial_{\sigma} \mathcal{W}_{eff}=0.
	\label{dW=0}
\end{align}
Let us first show that the roots of this equation  (VEVs of $\sigma$) coincide with the roots of the Seiberg-Witten curve.  
The trace of the resolvent of the operator $\sqrt{2}\Phi$ can be expressed through the derivative:
\begin{align}
	\Tr\frac{1}{x-\sqrt{2}\Phi}
	=\frac{d}{dx}\log\det(\Sigma-\sqrt{2}\Phi) \,.
\end{align}
According to \cite{Cachazo:2003yc}, we find
\begin{align}
	& \int^{\Sigma} dx\,
	\left\langle  \Tr\frac{1}{x-\sqrt{2}\Phi}\right\rangle
	=\log\left(\frac{\sqrt{2}^N\left( P\left(\frac{\Sigma}{\sqrt{2}}\right)+y\left(\frac{\Sigma}{2}\right)\right)}{2} \right) \,,
\end{align}
and therefore \eqref{dW_1} takes the form 
\begin{align}
	4\pi\partial_{\Sigma}\mathcal{W}_{eff}(\Sigma)
	= \log\left(\frac{\left[P\left(\frac{\Sigma}{\sqrt{2}}\right)+y\left(\frac{\Sigma}{\sqrt{2}}\right)\right]^2}{4\left(\frac{\Lambda_{4D}}{\sqrt{2}}\right)^{2 N-N_f}Q(\frac{\Sigma}{\sqrt{2}})} \right) \,. 
\label{DW}
\end{align}
This is the general form of the twisted superpotential derived in \cite{Gaiotto:2013sma}, which applies to a broad class of surface defects whose world-sheet theories are $\mathbb{WCP}$ models. However, in the particular case where the two-dimensional theory is described at the classical level by the action $S_{(2,2)}^{\rm cl}$ introduced in Sec.~\ref{sec:r<N}, the superpotential has the following special property. Suppose that $\sigma_{P}$ corresponds to a vacuum of the two-dimensional theory, so the l.h.s. of \eqref{DW} at $\sigma_P$ is zero, and that $\sigma_{P}\neq m_K$ for all $K$. Then, from~\eqref{DW}, we obtain
\(
y\left(\frac{\sigma_{P}}{\sqrt{2}}\right)=0
\). Hence we arrive at the important conclusion
\begin{align}
	\sigma_P=\sqrt{2}e_P ,
	\label{equalroots_r}
\end{align}
which generalizes the relation \eqref{equalroots} to all $r$ vacua. For the $r=N-2$ vacuum, the roots of the Seiberg-Witten curve are calculated in Appendix \ref{append:roots_and_vevs} and this relation
is checked explicitly.
Thus, all vacua of the 2D theory coincide%
\footnote{We note that, while generically one can consider a large class of surface defects supporting various degrees of freedom, the dynamical vortex strings here are very specific. In particular, the 2D-4D correspondence is not at all a generic phenomenon for surface defects, but here it arises naturally for the confining string. 
}
(up to the factor $\sqrt{2}$) with the roots of the Seiberg--Witten curve in 4D.

Now let us use the representation \eqref{dW} to find the equation for $\sigma$ VEVs.
Technically, it is more convenient to work with an exponentiated version of the vacuum equation.
Substituting \eqref{dW} into \eqref{dW=0} we have:
\begin{align}
	t^{\left(2 N-N_f\right)} \prod_{P=1}^{r} \frac{\left(\sigma+m_P\right)}{t_P}=\Lambda_{4D}^{\left(2 N-N_f\right)} \prod_{K=r+1}^{N_f} \frac{\left(\sigma+m_K\right)}{t_K}, 
	\label{chiralring}
\end{align}
The quantities $t$ and $t_A$ are defined in Eq.~\eqref{t_t_A}. 
Eq.~\eqref{chiralring} is also known as the chiral ring equation.

\subsubsection{Quasiclassical limit and large roots}

Although the equation \eqref{chiralring} is exact, it cannot be solved analytically at arbitrary values of the mass parameters. Therefore, below we 
consider the quasiclassical large-mass limit $|m_A|\gg\Lambda_{4D}$.

The classical analysis of Sec.~\ref{sec:r<N} (see \eqref{classical_n})  shows that in the large-mass limit Eq.~\eqref{chiralring} should have  $ r$  large roots
\begin{align}
	\sigma_{P_0}\approx-m_{P_0}; \ P_0=1,\ldots r. 
	\label{lead_order}
\end{align}
The presence of large $O(m_P)$ roots can be easily checked   in the leading-order approximation, where  the gaugino condensate is small and \( \sigma \approx t \approx t_A \).
In this limit, Eq.~\eqref{chiralring} simplifies to
\begin{align}
	\sigma^{2\left(N-r\right)} \prod_{P=1}^{r} \left(\sigma+m_P\right)=\Lambda_{4D}^{\left(2 N-N_f\right)} \prod_{K=r+1}^{N_f}\left(\sigma+m_K\right).
	\label{chiral_ring_large-m}
\end{align}
In the limit \( \Lambda_{4D} \to 0 \), we recover \eqref{lead_order}. 
To find corrections to the roots $\sigma_{P_0}$, let us expand Eq.~\eqref{chiral_ring_large-m} around $\sigma \approx -m_{P_0}$:
\begin{align}
	m_{P_0}^{2\left(N-r\right)}(\sigma+m_{P_0}) \prod_{P\neq P_0}^{r} \left(-m_{P_0}+m_P\right)=\Lambda_{4D}^{\left(2 N-N_f\right)} \prod_{K=r+1}^{N_f}\left(-m_{P_0}+m_K\right),
\end{align}
This linearized equation is solved by
\begin{align}
	\sigma_{P_0} \approx-m_{P_0}+\frac{\Lambda_{4D}^{2N-N_f}\prod_{K=r+1}^{N_f}(m_K-m_{P_0})}{m_{P_0}^{2(N-r)}\prod_{P\neq P_0}^{r}(m_P-m_{P_0})};\quad P_0=1,\ldots r. \label{large_sigma}
\end{align}
If we assume that all masses are of the same order $m_A \sim m$, we see that the corrections are suppressed by powers $(\Lambda_{4D}/m)^{2N-N_f}$.

\subsubsection{Small roots and the gaugino condensate}
\label{sec:small_and_S}

Next, let us compute  the small roots $\sigma \sim 0$.
To do this, we  expand Eq.~\eqref{chiralring}  to leading order.
For small $\sigma \ll m_A$ and  $t\approx t_A$ (this will be justified below Eq.~\eqref{small_sigma}) 
Eq.~\eqref{chiralring} simplifies to
\begin{align}
	t^{2(N-r)}\approx \Lambda_{YM}^{2(N-r)}, \qquad \Lambda_{YM}^{2(N-r)} 
	\equiv \Lambda^{2N-N_f}_{4D} \frac{\prod_{K=r+1}^{N_f}m_K}{\prod_{P=1}^{r}m_P},
	\label{sigma-small_chiral_ring}
\end{align}
where we introduced the scale of 4D \ntwo pure Yang-Mills theory with the gauge group $U(N-r)$,
which has the first coefficient of the $\beta$-function $b_{YM}=2(N-r)$. This theory emerges in our 4D SQCD at energies below the quark mass scale, see Sec.~\ref{sec:r<N}.

Solving this equation for \(t\), we find $2(N-r)$ complex roots $t = t_k$ with  
\begin{align}
	t_k\approx e^{\frac{i\pi k}{N-r}}\, \Lambda_{YM}; \quad k=0,\ldots, 2(N-r)-1.
	\label{t_small_root_approximate_solution}
\end{align}
Then, using the definition of $t$ from Eq.~\eqref{t_t_A}, we arrive at  
\begin{align}
	\sigma_{k}=\frac{\frac{S}{\mu}+t_k^2}{t_k}; \quad k=0,\ldots, 2(N-r)-1.
	\label{sigma_small_root_approximate_solution}
\end{align}
These roots give the vacua of the two-dimensional theory in terms of the gaugino condensate $S$.
We now turn to determining its value.
Let us denote
\begin{align}
	\frac{4S}{\mu} \equiv a^2 .
\end{align}
Then the variables $t_A,t$ can be rewritten as
\begin{align}
	2t_A
	=
	\sqrt{\sigma^2-a^2}
	+
	\frac{m_A\sigma+a^2}{\sqrt{m_A^2-a^2}},
	\qquad
	2t
	=
	\sqrt{\sigma^2-a^2}
	+
	\sigma .
\end{align}
In addition to the variables $t_A$ and $t$, one can introduce the ``dual'' variables $\widetilde t_A$ and $\widetilde t$:
\begin{align}
	2\widetilde t_A
	=
	-\sqrt{\sigma^2-a^2}
	+
	\frac{m_A\sigma+a^2}{\sqrt{m_A^2-a^2}},
	\qquad
	2\widetilde t
	=
	-\sqrt{\sigma^2-a^2}
	+
	\sigma .
\end{align}
They satisfy
\begin{align}
	t_A \widetilde t_A
	=
	\frac{a^2(\sigma+m_A)^2}{4(m_A^2-a^2)},
	\qquad
	t\widetilde t
	=
	\frac{a^2}{4}.
\end{align}
Therefore, the equation \eqref{chiralring} can be rewritten in the dual form:
\begin{align}
	\widetilde t^{\left(2 N-N_f\right)}
	\prod_{P=1}^{r}
	\frac{\left(\sigma+m_P\right)}{\widetilde t_P}
	=
	\widetilde\Lambda_{4D}^{\left(2 N-N_f\right)}
	\prod_{K=r+1}^{N_f}
	\frac{\left(\sigma+m_K\right)}{\widetilde t_K},
	\label{chiralringdual}
\end{align}
where
\begin{align}
	\widetilde\Lambda_{4D}^{2 N-N_f}
	=
	\left(\frac{a^2}{4\Lambda_{4D}}\right)^{2N-N_f}
	\prod_{P=1}^r
	\frac{4(m_P^2-a^2)}{a^2}
	\prod_{K=r+1}^{N_f}
	\frac{a^2}{4(m_K^2-a^2)} .
	\label{lambda_dual}
\end{align}
One can analytically continue $\sigma$ around either $a$ or $-a$. This transformation acts as
\begin{align}
	\sigma \to \sigma,
	\qquad
	t \to \widetilde t,
	\qquad
	t_A \to \widetilde t_A,
\end{align}
without producing any additional multiplicative factors. Applying this transformation directly to \eqref{chiralring} and then comparing the result with \eqref{chiralringdual}, we obtain
\begin{align}
	\Lambda_{4D}^{2N-N_f}
	=
	\widetilde\Lambda_{4D}^{2 N-N_f}.
\end{align}
This condition imposes a nontrivial constraint on the gaugino condensate. From \eqref{lambda_dual} it follows that
\begin{align}
	\left(\frac{a^2}{4}\right)^{2N-N_f}
	\prod_{P=1}^{r}
	\frac{4(m_P^2-a^2)^2}{a^2}
	=
	\Lambda_{4D}^{4N-2N_f}
	\prod_{K=r+1}^{N_f}
	\frac{4(m_K^2-a^2)^2}{a^2}.
	\label{self_duality}
\end{align}
Let us note that this self-consistency condition also follows directly from the existence of two roots with the property
\begin{align}
	(\sigma^\pm_{N})^2\equiv\frac{4S}{\mu}. 
	\label{sigmaN_special_roots}
\end{align}
Solving \eqref{self_duality} in the large-mass limit, we determine the value of the gaugino condensate:
\begin{align}
	\frac{S}{\mu}\approx e^{\frac{2i\pi s}{N-r}}\, \Lambda_{YM}^2; \quad s=0,\ldots, N-r-1,
	\label{S}
\end{align}
This gives \(N - r\) discrete values for the gaugino condensate, as expected.
Each value corresponds to a different monopole/dyon vacuum. Formula \eqref{S} generalizes the known result \cite{Shifman:2014lba}  for the case \( r = N-1 \) and reproduces the answer for $r=N-2$ obtained by direct calculation in Appendix~\ref{append:roots_and_vevs}.

\begin{figure}[t]
	\centering
	\includegraphics[width=0.8\textwidth]{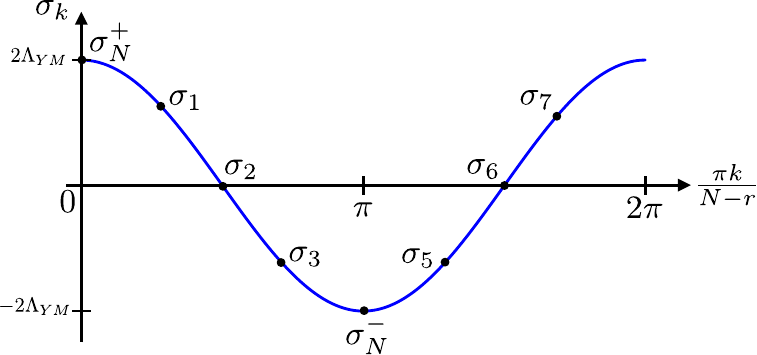} 
	\caption{Values of $\sigma_k$ at $s=0$ and $N-r=4$. Here, $\sigma_N^{+}=\sigma_0$ and $\sigma_N^{-}=\sigma_4$, while the remaining roots are equal in pairs: $\sigma_1=\sigma_7$, $\sigma_3=\sigma_5$ and $\sigma_2=\sigma_6$.}
	\label{fig:cos}
\end{figure}

Substituting the gaugino condensate \eqref{S} into Eq.~\eqref{sigma_small_root_approximate_solution} and using Eq.~\eqref{t_small_root_approximate_solution}, we finally arrive at
\begin{equation}
	\sigma_k \approx  2\cos\left(\pi\frac{k-s}{N-r}\right)e^{\frac{i\pi s}{N-r}}\,\Lambda_{YM},
	\qquad  k = 0,\ldots,2(N-r)-1  \,.
	\label{small_sigma}
\end{equation}
Here, $s$ is an integer parameterizing the phase of the gaugino condensate \eqref{S}, and it is held fixed, so we consider a given vacuum in 4D SQCD.  Let us fix it to be the monopole vacuum with
$s=0$ .
With this solution at hand, one can easily check that the approximation \( t \approx t_A \) holds for these roots, ensuring the self-consistency.
Note that the root $\sigma_k$ with $k=s$ gives $\sigma^+$, while $\sigma_k$ with $k=N+s-r$ gives $\sigma^-$ satisfying Eq.~\eqref{sigmaN_special_roots}.

The  set in \eqref{small_sigma} is overcomplete; see Fig.~\ref{fig:cos}, with some roots appearing twice. 
Let us summarize the $\sigma$-vacua of the world-sheet theory as follows:
\begin{itemize}
	\item The first $r$ large  $\sigma$ VEVs \eqref{large_sigma} correspond to magnetic strings.
	\item The two roots from \eqref{sigmaN_special_roots}.  We denote these VEVs  $\sigma_{N}^{\pm}$
	to make contact with two first-order roots $e_N^{\pm}$ of the Seiberg-Witten curve, see 
	\eqref{SW_curve_rep-01}. They should be identified with the ends of the open string, the ``strings'', which are absent in all $r<N$ 4D vacua in contrast to the $r=N$ vacuum, where  $N$ magnetic strings are present, see \cite{Shifman:2014lba}.
	\item Another $2(N-r-1)$ values $t_k$ in \eqref{t_small_root_approximate_solution}, in fact, correspond to only
	$(N-r-1)$ roots of $\sigma$ in Eq.~\eqref{chiralring}. In \eqref{small_sigma} they correspond to distinct values of $\sigma$. These $(N-r-1)$ vacua of the 2D theory  are associated with  electric strings.
\end{itemize}
We note that the results \eqref{large_sigma},\eqref{small_sigma} reproduce the known outcomes for the case \( r = N \), as well as for \( r = N-1 \) \cite{Shifman:2014lba}.

\section{2D-4D correspondence}
\label{2D_4D_corr}

The discrete set of vacua in the 2D world-sheet theory leads to the presence of kinks interpolating between different 2D vacua. In this section we will find the masses of  these kinks
and show that their masses are equal to the masses of confined monopoles or quarks, generalizing the 2D-4D correspondence found earlier for
$r=N$ and  $r=N-1$ 4D vacua; see \cite{Shifman:2014lba} and the review \cite{SYrev}.

\subsection{Kink masses in 2D}
\label{sec:kinkmasses}

The  central charge  of a BPS kink interpolating between the
vacua $\sigma'$ and $\sigma''$ is given by the 
difference of the exact superpotential $\mathcal{W}_{eff}$ calculated at distinct roots of the chiral ring equation  \cite{Dorey,DoHoTo,HaHo}.  Explicitly, for the kink mass we have 
\begin{align}
	M^{\text{kink}}= 2 | \mathcal{W}(\sigma'')-\mathcal{W}(\sigma')|=2 \  \Bigg|\int_{\sigma'}^{\sigma''} \partial_{\sigma}\mathcal{W}d\sigma
	\Bigg|=2 \Bigg|\int_{\sigma'}^{\sigma''} \sigma\partial_{\sigma}^2\mathcal{W} d\sigma
	\Bigg|. \label{6.1}
\end{align}
Eq.\eqref{d^2W} leads us to
\begin{align}
	M^{\text{kink}}=\Bigg| \frac{1}{2\pi}\int^{\sigma''}_{\sigma'} \Sigma  d\Sigma\left\{2 \left\langle \Tr\frac{1}{\Sigma-\sqrt{2}\Phi} \right\rangle-\sum_{A=1}^{N_f}\frac{1}{\Sigma+m_A}
	\right\}\Bigg| \,.
\label{6.2}
\end{align}
Let us rewrite this formula in a more explicit fashion using the particular expression Eq.~\eqref{exact_resolvent} for the resolvent  
\begin{align}
	M^{\text{kink}} =\left|\frac{1}{2\pi }\int^{\sigma''}_{\sigma'}\frac{\sigma d\sigma}{\sqrt{\sigma^2-\frac{4S}{\mu}}}\left[2N-N_f -\sum_{j=1}^{r}\frac{\sqrt{m_j^2-\frac{4S}{\mu}}}{\sigma+m_j}+\sum_{j=r+1}^{N_f}\frac{\sqrt{m_j^2-\frac{4S}{\mu}}}{\sigma+m_j}\right] \right|,
	\label{mkink}
\end{align}

In the next subsection we will calculate the masses of massive monopoles and quarks  at the singular point
on the  Coulomb branch where $r$ quarks and $(N-r-1)$ monopoles are massless. This point becomes an isolated  $r$ vacuum once we switch on small $\mu$. We will see that  the formulas \eqref{SW_quark_mass_period} and \eqref{monopoles_masses} for massive monopoles/quarks identically coincide with the one in \eqref{mkink} or \eqref{6.2}. This proves our generalization of the 2D-4D correspondence to all $r$ vacua
of \ntwo SQCD.

The integral in \eqref{mkink} can be straightforwardly evaluated, giving the final result
\begin{equation}
	\begin{aligned}
		& 
		M^{\text{kink}} =\left|\frac{1}{2\pi }\left\{(2N-N_f)\sqrt{\sigma^2-\frac{4 S}{\mu}}
		- \sum_{A=1}^{r} \left[\sqrt{m_A^2-\frac{4S}{\mu}}\log{t }+m_A
		\log{\left(\frac{\sigma+m_A}{t_A} \right)}
		\right] \right.\right.
		\\[3mm]
		&\left.\left.
		+ \sum_{A=r+1}^{N_f} \left[\sqrt{m_A^2-\frac{4S}{\mu}}\log{t}+m_A
		\log{\left(\frac{\sigma+m_A}{t_A} \right)}
		\right]\right\}\right|_{\sigma'}^{\sigma''} \,.
	\end{aligned}
	\label{mkink_fin}
\end{equation}
Note also that, depending on a particular choice of the integration contour in Eq.~\eqref{mkink}, the integral can pick up the residues located at $\sigma = - m_j$.
Such a residue would give a contribution of $i \, m_j$ times an integer, which corresponds to dyonic kinks. 

To illustrate this, consider the simplest example of the $r=1$ vacuum in SQCD with $N=N_f=3$, see
Sec.~\ref{sec:example_U3_r1}. In the large $m_A$ limit we have in the  world-sheet theory one  large $\sigma$-vacuum
\beq
\sigma_1\approx-m_1+\Lambda_{4D}^3 \frac{(m_2-m_3)(m_3-m_1)}{m_1^5} \,,
\label{sigma_1}
\eeq
associated with a magnetic string. The equation \eqref{sigma-small_chiral_ring} for small roots and the result \eqref{S} for the gaugino condensate take the form
\begin{align}
	t^{4}\approx \Lambda_{YM}^{4}, \qquad \Lambda_{YM}^{4} 
	\equiv \Lambda^{3}_{4D} \frac{m_2m_3}{m_1}, \qquad \frac{S}{\mu}\approx \Lambda_{YM}^2,
	\label{t_eq}
\end{align}
where we focus on the  monopole vacuum $s=0$. This equation has three roots,
\beq
\sigma_2\approx 0, \qquad \sigma^{\pm}_3\approx\pm 2\Lambda_{YM}.
\label{small_sigmas}
\eeq

Let us find the mass of the kink interpolating between the 2D $\sigma_1$  vacuum and endpoint roots
$\sigma_3^{\pm}$. Eq.~\eqref{mkink_fin} gives
\beq
\begin{aligned}
	M^{kink}_{\sigma_1 \sigma_{3}^{\pm}}&\approx \\
	&\frac{1}{2 \pi} \Bigg|3 m_1 \log\left(\frac{m_1}{\Lambda_{4D}}\right)+ \sum_{A=2,3}\left[(m_A-m_1)\log\left(\frac{m_A-m_1}{m_1}\right)- m_A\log\left( \frac{m_A}{m_1}\right)\right]\Bigg|
	\label{m-kink_mass},
\end{aligned}
\eeq
which does not depend on the choice of final root $\sigma_3^{+}$ or $\sigma_3^{-}$ \cite{Shifman:2014lba}. This quantum result matches the quasiclassical expression for the monopole mass \eqref{mon_mass_cl}, where we should put $m_r=m_1$. To see this, note that the leading term in the $m_A\gg \Lambda_{4D}$ limit proportional to $\log{\Lambda_{4D}}$ in \eqref{m-kink_mass}  comes with the correct coefficient $b/2\pi = 3/2\pi$.

The $\sigma$-vacuum $\sigma_2=0$ in \eqref{small_sigmas} is identified with an electric string.
It should confine quarks $q^{2A}$ and $q^{3A}$,  see Sec.~\ref{sec:example_U3_r1}. Eq.~\eqref{mkink_fin}   
gives the following masses for the corresponding kinks
\beq
\begin{aligned}
	M^{kink}_{\sigma_2 \sigma_{3}^{-}}&\approx\left|
	\frac{4}{\pi}\Lambda_{YM}+ 
	m_A\right|,  \\
	M^{kink}_{\sigma_{3}^{+}\sigma_2 }&\approx\left|-
	\frac{4}{\pi}\Lambda_{YM}+ 
	m_A\right|,\\
	A&=1,2,3,
\end{aligned}
\label{q-kink_mass}
\eeq
where we add particular dyonic terms  to small  quantum corrections $\sim \Lambda_{YM}$ for consistency of our picture. 

\begin{figure}[t]
	\centering
	\includegraphics[width=1\textwidth]{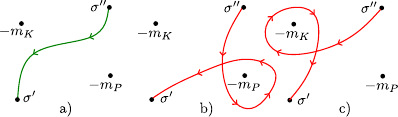} 
	\caption{Possible integration contours. The green contour determines the mass of monopole kinks, while the red contours give the masses of quark kinks. The orientation of the latter is fixed by the flavor: here $K = 1,\ldots,r$ and $P = r+1,\ldots,N_f$.  }
	\label{fig:dyonic_windings}
\end{figure}

The general prescription is as follows. For monopole kinks, the contour must not encircle any of the points $-m_j$.
For kinks corresponding to quarks, the contour is required to encircle
the pole at $-m_A$ exactly once, in accordance with the flavor label $A$.
The orientation of this encircling depends on whether $A \leq r$ or $A > r$; see  Fig.~\ref{fig:dyonic_windings}.

We stress that the masses of kinks in \eqref{m-kink_mass} and \eqref{q-kink_mass} coincide with the masses of the corresponding monopoles and quarks in the 4D theory, see Sec.~\ref{sec:masses_of_states} below.

Note also that the splitting of the masses of kinks associated with confined quarks of different colors in \eqref{q-kink_mass} is due to the breaking of the gauge group $U(N-r)$ down to $U(1)^{N-r}$ by condensation of adjoint scalars at the quantum level.

To conclude this subsection, let us discuss the correction  $S_{def}$ proportional to $\mu$ to the
world-sheet action. The obvious generalization of the world-sheet potential \eqref{S_def} to $r$ vacua was suggested in \cite{Shifman:2014lba}. It has the form
\begin{align}
	V_{def}^{2D} = 4 \pi  \left|\mu\sqrt{\sigma^2-\frac{4S}{\mu}}\right|. 
	\label{S_def_r}
\end{align}
This potential correctly reproduces string tensions $T_P=2\pi\xi_P$ with $\xi_P$ given by \eqref{tension}  in the 2D
$\sigma$-vacua determined by roots of the chiral ring equation \eqref{chiralring} due to the correspondence \eqref{equalroots}. At two points, $\sigma_N^{\pm}$,  the vacuum energy is zero.
This vacuum corresponds to the nonexistent $N$th string.

Note that tensions of $(N-r-1)$ electric strings of the $U(N-r)$ Yang-Mills sector of our SQCD are given by substitution of small roots from \eqref{small_sigma} to \eqref{S_def_r}. This  gives
\beq
T_k \approx 8\pi  \left|  \mu \sin{\left(\frac{\pi k}{N-r}\right) }\Lambda_{YM}\right|, \qquad k=1,\ldots (N-r-1),
\label{sin_eq}
\eeq
which is the  famous sine formula for the string tensions in the monopole vacuum of Seiberg-Witten theory \cite{DougShenk}, see also \cite{HanStrassZaf,Shifman:2013zsa}. Notably, together with higher windings \cite{VY,Strassler:1997ny}, this can lead to extra states in the spectrum compared to QCD.

The  deformation potential \eqref{S_def_r}
breaks \ntwot  2D supersymmetry down to $\mathcal{N}=(0,2)$, 
which is further spontaneously broken by choosing a vacuum with nonvanishing energy 
\cite{Shifman:2014lba}.

Strictly speaking, at finite $\mu$ the potential \eqref{S_def_r}  renders the kinks of the 2D theory metastable.
Nevertheless, this effect disappears in the limit $\mu \to 0$, so below we consider the kinks to be stable.

\subsection{Masses of 4D confined constituents}
\label{sec:masses_of_states}

Let us summarize the confinement mechanism for an arbitrary \( r \) vacuum, see Appendix~\ref{sec:appendix_charges} for a more detailed discussion. The monopoles
\( M_{r+1,r+2}, \ldots, M_{N-1,N} \) and the squarks
\( q^{11}, \ldots, q^{rr} \) are condensed. Each of them serves as a source of a chromo-electric (magnetic) string. The elementary monopoles \( M_{1,2}, \ldots, M_{r-1,r} \) correspond to junctions of neighboring magnetic strings, while the monopole \( M_{r,r+1} \) represents an endpoint of the string.
Similarly, the quarks \( q^{N-1\,A} \) and \( q^{N\,A} \) are attached to the endpoints of electric strings, whereas \( q^{r+1\,A}, \ldots, q^{N-2\,A} \) correspond to string junctions. Both quarks and monopoles form necklace-type mesons, as in the cases discussed in Secs.~\ref{sec:mesonic_1} and~\ref{U_5_r_2}.

\begin{figure}[h!]
	\centering
	\includegraphics[width=1\textwidth]{
		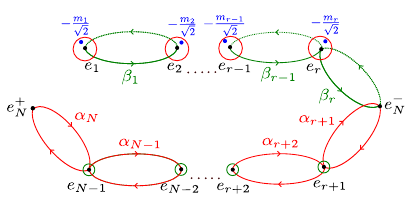} 
	\caption{Black dots are roots of Seiberg-Witten curve. Red curves represent the $\alpha$ cycles, while the green ones are the $\beta$ cycles. Dashed lines lie on the second sheet. Blue dots label the points with non-trivial residue of Seiberg-Witten  differential. The number of non-vanishing $\alpha_i$-cycles coincides with the number of confined quark colors $q^{kA}$. The indices of the red contours $\alpha_i$ denote the corresponding unique position of the quark states $q^{kA}$ in ``necklace" configuration. The indices of the green contours $\beta_j$ refer to the monopoles $M_{j,j+1}$.}
	\label{fig:roots}
\end{figure}

Having identified  the quarks and monopoles that become confined, let us now turn to a more detailed 4D description of these constituents.
In the present supersymmetric setup, these particles come in BPS supermultiplets.
As such, their masses can be found from the Seiberg-Witten solution.

At a generic point of the Coulomb branch, the masses of the BPS states are determined by $\alpha$ and $\beta$ periods of the curve \eqref{swcurve}.  We are interested in the singular points on the Coulomb branch, which become $r$-vacua upon switching on the small $\mu$-deformation, see  \eqref{def}. The Seiberg-Witten curve degenerates to the one in Eq.~\eqref{SW_curve_rep-01} with $r$ double roots associated with massless quarks and $N-r-1$ double roots associated with massless monopoles. This corresponds to the shrinking
of $r$ $\alpha$-cycles and $(N-r-1)$ $\beta$-cycles, respectively, see Fig~\ref{fig:roots}.

Masses of the elementary monopoles on the Coulomb branch in the would-be $r$-vacuum (a singular point which becomes the $r$-vacuum at nonzero $\mu$) are then given by
\begin{equation}
	\begin{aligned}
		M^{\text{monopole}}_{jj+1} &= \left| \frac{\sqrt{2}}{2\pi i}\oint_{\beta_{j}}\lambda_{\mathrm{sw}}\right| \,, \quad j=1,\ldots,r \\
		M_{ii+1}^{\text{monopole}} &= 0 \,, \quad i=r+1,\ldots,N-1,
	\end{aligned}
	\label{monopoles_masses}
\end{equation}
where $\lambda_{\mathrm{SW}}$ is the Seiberg--Witten differential \cite{SW,SW2,HaOz,GMMM}, which is related to the resolvent \eqref{resolvent} as follows:
\begin{align}
	\lambda_{SW}=x\left\langle \Tr\frac{1}{x-\Phi} \right\rangle dx \,.
\end{align}

The vanishing monopole masses correspond to the shrinking $\beta$-cycles of the SW curve, see Fig~\ref{fig:roots}.
Note that the masses of the junction-like monopoles are evaluated by the periods connecting two quark double roots, while the endpoint monopoles correspond to the periods connecting a quark double root with one of the  simple roots  \( e_N^{\pm} \).
This allows us to identify these roots with the left and right endpoints of the monopole necklace configuration.

Now let us turn our attention to the quark hypermultiplets.
Their masses are given by a general formula
\begin{align}
	m^{q}_{k,A}= \left| \frac{\sqrt{2}}{2\pi i}\oint_{\alpha_{i}}\lambda_{\mathrm{sw}}+m_A\right|.
	\label{SW_quark_mass_period}
\end{align}
Let us take a closer look at the condensed squarks. 
As their BPS masses vanish, we demand that
\begin{align}
	\underset{x=-\frac{m_P}{\sqrt{2}}}{\mathrm{Res}}\lambda_{\mathrm{sw}}=-\frac{m_P}{\sqrt{2}}, \quad P=1,\ldots r, 
	\label{res_masses}
\end{align}
which guarantees that Eq.~\eqref{SW_quark_mass_period} yields zero for $P=A$.
We further suppose that $\lambda_{\mathrm{sw}}$ is holomorphic at the other bare masses.
The condition \eqref{res_masses} allows us to further specify the quark mass formula:
\begin{equation}
	\begin{aligned}
		m^{q}_{k,A} &= \left| \frac{\sqrt{2}}{2\pi i}\oint_{\alpha_{i}}\lambda_{\mathrm{sw}}+m_A\right| \,; \quad i,k \in \{r+1,\ldots,N\} \\
		m^{q}_{p,A} &= |m_{p}-m_A| \,, \quad p=1,\ldots,r \\
		A & =1,\ldots,N_f
	\end{aligned}
	\label{quark_masses_general}
\end{equation}
As with the monopoles, the masses of the quarks sitting at the endpoints of an electric string, namely $m^q_{N-1, A}$ and $m^q_{N A}$, are determined by contours $\alpha_{r+1}$ and $\alpha_N$ passing through the simple roots $e_{N}^{\pm}$.
This is most easily seen in the $r=N-2$ case, see Sec.~\ref{sec:example_U3_r1}.
For each value of the quark's color label $k$, there is a unique contour $\alpha_i$. The precise correspondence between $k$ and $i$ (apart from the just discussed $k=N-1,N$) is not very important for the purposes of this work. Two simple examples of exact matching are presented in Appendix~\ref{append:n3}.

Contour integrals in Eqs.~\eqref{monopoles_masses} and \eqref{quark_masses_general} can be reduced in the following way:
\begin{equation}
	\begin{aligned}
		\frac{\sqrt{2}}{2 \pi i}\oint_{\alpha_i;\beta_j}\lambda_{\text{SW}} &= \frac{\sqrt{2}}{\pi i}\int_{e'}^{e''}\left[\frac{x dP(x)}{y(x)}- \frac{xP(x)dQ(x)}{2y(x)Q(x)}\right]\\
		&= \frac{\sqrt{2}}{\pi i}\int_{e'}^{e''}x\left\{2 \left\langle \Tr\frac{1}{x-\Phi} \right\rangle-\sum_{A=1}^{N_f}\frac{1}{x+\frac{m_A}{\sqrt{2}}}
		\right\}dx \,.
	\end{aligned}
	\label{SW_periods_key_formula}
\end{equation}
Here, the integration limits $e'$ and $e''$ should be chosen as the appropriate SW roots for a given contour  $\alpha_i$ and $\beta_j$.   
Changing variables of integration $\sqrt{2}x=\sigma$ and taking into account the coincidence of the limits Eq. \eqref{equalroots}, we see that the answer matches precisely the mass formula for the 2D kinks \eqref{6.2},   which was derived from the 2D exact superpotential \eqref{DW}.
This is, of course, a manifestation of the 2D-4D correspondence. 

We now identify which kinks correspond to confined monopoles and quarks.
As discussed in Sec.~\ref{sec:confiniment}, each confined state is in one-to-one
correspondence with a confining string or with a pair of confining strings.
Each such string is uniquely associated with an isolated vacuum of
$S^{\mathrm{eff}}_{(2,2)}$ and with a root of the Seiberg--Witten curve.
Accordingly, for each state one can unambiguously specify the corresponding
$\alpha$- or $\beta$-cycle, as well as a pair of roots $(e'', e')$,
which determine the mass of the confined monopole or quark. In the two-dimensional description, for a given state one must specify
an integration contour in the $\sigma$-plane, starting at
$\sigma'=\sqrt{2}\,e'$ and ending at $\sigma''=\sqrt{2}\,e''$,
see  Fig.~\ref{fig:dyonic_windings}.

\section{Conclusions}

\label{sec:concl}

In this paper, we constructed the world-sheet theory for the non-Abelian string in the $r$ vacuum of 
\ntwo SQCD for arbitrary $r\le  (N-2)$. For  the $r=N$ quark vacuum, the world-sheet theory is given by the 
2D \wcpNt model; see \cite{SYrev} for a review. The case $r=N-1$ was studied in \cite{Shifman:2014lba}. In this case, the exact twisted superpotential is affected by  instantons of 4D SQCD,
generating the gaugino condensate. The new feature of  the hybrid $r<N-1$ vacua considered in this paper is the presence of $r$ condensed quarks and $(N-r-1)$ condensed monopoles. As a result, both color-magnetic and color-electric strings are formed, which confine monopoles and quarks, respectively. 

Using the  Gaiotto–Gukov–Seiberg method of resolvents \cite{Gaiotto:2013sma}, we constructed the exact twisted superpotential of the world-sheet theory on the non-Abelian string, whose 2D vacua correspond to both elementary magnetic and electric strings. This provides a unified description of both types of strings. 

In particular, this means that Abrikosov-Nielsen-Olesen Abelian strings  formed due to the monopole condensation and   responsible for the confinement of quarks in the Seiberg-Witten scenario are seen as particular vacua in our world-sheet theory on the non-Abelian string.

Next, using this exact superpotential, we calculated masses of kinks interpolating between 2D vacua
and showed that they coincide with the masses of monopoles and quarks on the Coulomb branch at the  singular point corresponding to the $r$-vacuum. The explanation of this coincidence is that confined monopoles and quarks of 4D SQCD are seen as kinks
in the world-sheet theory on the non-Abelian string.  This result generalizes the 2D-4D correspondence found earlier for $r=N$ and $r=N-1$ vacua.

\section*{Acknowledgments}

The work of D.V. was supported by the Ministry of Science and Higher Education of the Russian Federation (agreement 075-15-2025-344 dated 29/04/2025 for Saint Petersburg Leonhard Euler International Mathematical Institute at PDMI RAS). The work of E.I. is supported in part by U.S. Department of Energy Grant No. de-sc0011842.

\appendix

\section{Charges in the low-energy EFT}
\label{sec:appendix_charges}

To understand which monopoles are affected by quark VEVs and vice versa, it is useful to consider a generic scenario.
Let us start from a theory with a gauge group $G$.
On the Coulomb branch it is spontaneously broken to its maximal torus,
\begin{equation}
	G \to U(1)^{\text{rk}_G} \,.
	\label{appendix_G_breaking}
\end{equation}
Here, $\text{rk}_G$ is the rank of the gauge group.
Let us denote the Cartan generators as $H_a$, $a=1, \ldots, \text{rk}_G$, normalized with respect to a certain norm. For example, if the Lie algebra $\mathfrak{g}$ of $G$ is a semisimple algebra over $\mathbb{C}$, one can normalize the generators with respect to the Killing form as follows:
\begin{equation}
	\Tr (H_a H_b) = \frac{1}{2} \delta_{ab} \,.
	\label{appendix_H_normalization}
\end{equation}
This convention is compatible with Eq.~\eqref{algebra_normalization} employed in this paper.

\subsection{Weights, co-roots and charges}

Quark hypermultiplets are in the fundamental representation of $G$. 
On the Coulomb branch they acquire Abelian electric charges $ \mathtt{Q}^{\text{el}, (j)}_a = \vec{w}^{(j)}_a \equiv$ that are in the weight lattice,  $\vec{w}^{(j)} \in \Lambda_W$.
Here, the index $j$ enumerates different sets of electric charges.

This can be seen as follows.
A given squark has the $\Tr |A_\mu q^{(j)}|^2$ term in the action.
At low energies, we can write $A_\mu = \sum_a A_\mu^a H_a$ corresponding to the unbroken part of the gauge group.
A weight vector $w^{(j)}_a$ in a given (fundamental) representation can be understood as a set of eigenvalues of the Cartan elements, i.e.
\begin{equation}
	H_k q^{(j)} = w^{(j)}_k q^{(j)} \quad
	\text{ (fixed } j) \,.
	\label{G_weights_definition}
\end{equation}
Here, the squark $q^{(j)}$ is restricted to be the corresponding eigenvector, which typically amounts to taking a specific (i.e. $j^\text{th}$) color component.
With the definition \eqref{G_weights_definition} in mind, the $\Tr |A_\mu q^{(j)}|^2$ interaction term for this flavor becomes
\begin{equation}
	\Tr \abs{ \sum_a A_\mu^a H_a q^{(j)} }^2 = \sum_a (\mathtt{Q}^{\text{el}, (j)}_{a})^2 \abs{ A_\mu^a q^{(j)} }^2 \,.
	\label{appendix_quark_electric_charges_IR}
\end{equation}
For a concrete example, see Eq.~\eqref{appendix_suN_electric_charges} below.

Monopoles at low energies acquire magnetic charges with respect to the $U(1)$ factors in Eq.~\eqref{appendix_G_breaking}.
For a monopole flavor $j$, the set of magnetic charges $\mathtt{Q}^{\text{mag}, (j)}_a$, $a=1, \ldots, \text{rk}_G$, belongs to the co-root lattice,
\begin{equation}
	\mathtt{Q}^{\text{mag}, (j)} \in \Lambda_R^*
	\label{magnetic_charge_coroot}
\end{equation}
To understand this, let us start by writing down the Dirac quantization condition.
Consider a single monopole with some particular set of magnetic charges $\mathtt{Q}^\text{mag}_{a}$.
Any probe electric (weight-like) charge $\mathtt{Q}^\text{el}_{a}$ corresponds to a Wilson line $\exp( i \oint \sum_a\mathtt{Q}^\text{el}_{a} A_\mu^a dx^\mu )$ which must be single-valued.
The magnetic field of the monopole is $\vec{B}^a = \mathtt{Q}^\text{mag} \vec{x}/|\vec{x}|^3$, and it leads to the Dirac quantization condition\footnote{ The factor 1/2 on the right hand side comes from the normalization \eqref{appendix_H_normalization} common in physics. }
\begin{equation}
	\sum_a\mathtt{Q}^\text{el}_{a} \mathtt{Q}^\text{mag}_{a} \in \frac12 \,\mathbb{Z}
	\label{dirak_quant}
\end{equation}
for any electric charge that we might have in the theory. 
Because we may have any electric charge from the weight lattice, this demands that the magnetic charge vector be at least from the co-weight lattice.
In the simplest cases like $SU(N)$ this is enough (co-weight lattice and co-root lattice coincide); however, in the more general case we should put further restrictions, which shrink it to the co-root lattice anyway.

\subsection{VEVs}

When electric or magnetic charges develop VEVs, they break some of the $U(1)$'s from Eq.~\eqref{appendix_G_breaking}.

Suppose that a single quark $q$ with charges $\mathtt{Q}^\text{el}_{a}$ develops a VEV.
The corresponding combination of the gauge fields, $\sum_a\mathtt{Q}^\text{el}_{a} A_\mu^a $, becomes Higgsed.
Any other electric charges aligned with $\mathtt{Q}^\text{el}_{a}$ become screened.
Furthermore, magnetic charges $\mathtt{Q}^\text{mag}_{a}$ that have a non-zero scalar product with $\mathtt{Q}^\text{el}_{a}$ (in the sense of Eq.~\eqref{dirak_quant}) become confined.

When, instead, a magnetically charged particle $\mathtt{Q}^\text{mag}_{a}$ develops a VEV, it screens all other magnetic charges proportional to $\mathtt{Q}^\text{mag}_{a}$, while confining the corresponding electric charges.

\subsection{The case with $G=U(N)$}

Now, let us consider a particular example with gauge group $U(N) = U(1) \times SU(N) / \mathbb{Z}_N$ and $N_f$ fundamental quark flavors.
According to the Seiberg-Witten theory, the monopoles that may be relevant in the IR come from the non-Abelian part of the gauge group.
Because the rank of $SU(N)$ is $\text{rk}_{SU(N)} = N-1$, we have $N-1$ basis species of monopoles in this theory.
At a generic point of the $\mathcal{N}=2$ Coulomb branch, the gauge group is broken down to (cf. Eq.~\eqref{appendix_G_breaking})
\begin{equation}
	U(N) \to U(1)^N \,.
	\label{appendix_uN_breaking}
\end{equation}
Condensation of quarks and/or monopoles breaks this further down to a single $U(1)$; we will discuss the effects of this breaking.

In order to see the qualitative picture of confinement, it is useful to consider the following basis for Cartan generators of $U(N)$:
\begin{equation}
	(H_a)_{ij} = \frac{1}{\sqrt{2}} \delta_{ai} \delta_{aj} \,, \quad
	a = 1, \ldots, N \,.
	\label{suN_dumb_cartans}
\end{equation}
Here, we represent the algebra generators as $N \times N$ matrices in the color space.
The normalization follows Eq.~\eqref{appendix_H_normalization}.

Consider a quark component with a fixed color $k_0$ (each quark flavor has $N$ color components).
Such a quark can be represented by an $N$-component vector which has only one non-vanishing component in the $k_0^\text{ th}$ position.
Acting with each of the generators \eqref{suN_dumb_cartans} on this vector, we infer the weight and, consequently, the electric charges of this quark (cf. Eqs.~\eqref{G_weights_definition} and \eqref{appendix_quark_electric_charges_IR}):
\begin{equation}
	\mathtt{Q}^{\text{el}, (k_0)}_a = \frac{1}{\sqrt{2}} \delta_{k_0 \,a} \,, \quad
	a=1,\ldots,N \,, \quad
	k_0 = \text{ fixed color.}
	\label{appendix_suN_electric_charges}
\end{equation}

The co-roots, or possible magnetic charges, come only from the non-Abelian part.
They are determined by the co-roots of $SU(N)$, see Eq.~\eqref{magnetic_charge_coroot}.
For illustration, let us pick the following basis of elementary magnetic charges:
\begin{equation}
	\mathtt{Q}^{\text{mag}, (j)}_{a} = \frac{1}{\sqrt{2}} (\delta_{a,j} - \delta_{a,j+1}) \,, \quad
	a = 1,\ldots,N \,, \quad 
	j = 1, \ldots, N-1 \,.
	\label{appendix_suN_magnetic_charges}
\end{equation}
Here, the index $j$ enumerates different basis elements.
Each basis element is an $N$-component vector, with the components labeled by the index $a$.
As one can see, the charges in Eqs.~\eqref{appendix_suN_electric_charges} and \eqref{appendix_suN_magnetic_charges} follow the Dirac quantization condition \eqref{dirak_quant}.

\subsubsection{Single-flavor condensation}

As a warm-up, suppose that one flavor of quarks develops a VEV. 
As a result, two monopoles become confined, forming a composite object whose net magnetic charge remains free.

For example, if the condensed quark has electric charge $\mathtt{Q}^\text{el} = \frac{1}{\sqrt{2}} (1,0,\ldots,0)$, it confines a pair of monopoles with magnetic charges $\mathtt{Q}^{\text{mag},(1)} = \frac{1}{\sqrt{2}} (1,-1,0,\ldots,0)$ and $\mathtt{Q}^{\text{mag},(N)} = \frac{1}{\sqrt{2}} (-1,0,\ldots,0,1)$.
These two monopoles then form a composite object with the net magnetic charge $\mathtt{Q}^{\text{mag},(1N)}= \frac{1}{\sqrt{2}} (0,-1,\ldots,0,1)$, and such an object can exist in isolation (it has a finite energy).

Now, suppose instead that a monopole with a set of magnetic charges $\mathtt{Q}^\text{mag,(1)}= \frac{1}{\sqrt{2}} (1,-1,0,\ldots,0)$ develops a VEV.
In this situation, one particular combination of elementary electric charges $\mathtt{Q}^\text{el,(conf)}= \frac{1}{\sqrt{2}} (1,-1,0,\ldots,0)$ is confined, while $\mathtt{Q}^\text{el,(free)}= \frac{1}{\sqrt{2}} (1,1,0,\ldots,0)$ and all other elementary electric charges remain unconfined.

\subsubsection{Generic $r$}

Now let us consider the case when $2 \leqslant r \leqslant N-2$ flavors of quarks become massless and develop VEVs.
From the argument above, it follows that $r+1$ out of $N-1$ species of monopoles become confined.
Naively, there could be several scenarios depending on which of the quarks condense: all $r+1$ monopoles may be confined to form a single molecule, or they may form several disjoint configurations.
In order to determine which scenario is realized, further data is needed.

One additional input is provided by the SW theory at a small $\mathcal{N}=1$ preserving deformation: most of the Coulomb branch of the $\mathcal{N}=2$ theory is lifted, leaving only an isolated set of points.
In each supersymmetric vacuum, a number of \textit{mutually local} BPS states become massless and develop VEVs, cf. Fig.~\ref{fig:roots}.

Let us say that the $r$ condensed squarks have the following electric charges with respect to the $U(1)^{N}$ gauge groups from Eq.~\eqref{appendix_uN_breaking}:
\begin{equation}
	\mathtt{Q}^{\text{el}, (j)}_a = \frac{1}{\sqrt{2}} \delta_{aj} \,, \quad
	a = 1, \ldots, N \,, \quad
	j=1, \ldots, r \,.
	\label{condensed_squarks_charges_suN_dumb}
\end{equation}
In that case, $r$ out of the $N$ gauge groups $U(1)$ from Eq.~\eqref{appendix_uN_breaking} are broken.
The broken $U(1)$ are the ones along the Cartan generators $H_a$, $a=1, \ldots, r$, see Eq.~\eqref{suN_dumb_cartans}; there are $r$ possible magnetic strings (tubes) carrying corresponding fluxes.
As a result, $r+1$ species of monopoles become confined.
The ones with magnetic charges $\mathtt{Q}^{\text{mag}, (j)}_{a}$ with $j=1, \ldots, r-1$ (as defined in Eq.~\eqref{appendix_suN_magnetic_charges}) are attached to two different magnetic strings, one carrying a magnetic flux along the $H_j$ and the other along the $H_{j+1}$ Cartan generator.
Furthermore, the monopoles $\mathtt{Q}^{\text{mag},(j)}_{a}$ with $j=N-1$ and $j=N \simeq 0$ are special --- they can serve as the endpoints of the confining thread.
Examples of the resulting configurations are shown in Fig.~\ref{fig:zoo}.

In a supersymmetric vacuum, on the quantum level $N-r-1$ corresponding monopoles/dyons develop VEVs.
In accordance with Eq.~\eqref{condensed_squarks_charges_suN_dumb}, the condensed monopoles have elementary magnetic charges
\begin{equation}
	\mathtt{Q}^{\text{mag},(j)}_a = \frac{1}{\sqrt{2}} (\delta_{aj} - \delta_{a,j+1}) \,, \quad
	a = 1, \ldots, N \,, \quad
	j=r+1, \ldots, N-1 \,.
	\label{condensed_monopoles_charges_suN_dumb}
\end{equation}
The electric charges of condensed squarks, Eq.~\eqref{condensed_squarks_charges_suN_dumb}, do not overlap with the magnetic charges of condensed monopoles, Eq.~\eqref{condensed_monopoles_charges_suN_dumb}, which ensures their mutual locality\footnote{We do not consider possible Argyres-Douglas points.}
Repeating the argument above, we can show that there are $N-r-1$ possible types of electric strings.
The quarks with electric charges 
\begin{equation}
	\mathtt{Q}^{\text{el},(j)}_a = \frac{1}{\sqrt{2}} \delta_{aj} \,, \quad
	a = 1, \ldots, N \,, \quad
	j = r+1, \ldots, N \,.
	\label{confined_squarks_charges_suN_dumb}
\end{equation}
have non-zero overlap with the condensed magnetic charges \eqref{condensed_monopoles_charges_suN_dumb} (in the sense that the Dirac pairing \eqref{dirak_quant} is non-zero).
Therefore, these quarks are confined by electric flux tubes.
In Eq.~\eqref{confined_squarks_charges_suN_dumb}, each quark with $j = r+2, \ldots, N-1$ overlaps with charges of two elementary monopoles from Eq.~\eqref{condensed_monopoles_charges_suN_dumb}, and so each of those quarks has two flux tubes attached to it.
The quarks with $j = r+1$ and $j = N$ are attached to a single flux tube; they can serve as endpoints of the string.

\subsubsection{Edge cases}

In the above analysis, we assumed a generic $r$ from an interval $ 2 \leqslant r \leqslant N-2$.
The two remaining cases are special.

When $r = N-1$, there is only a single species of monopoles (together with the corresponding antiparticle) that can sit at the end of a string, see  \cite{SYrvacua}.
At $r=N$, each monopole is necessarily a junction of two different strings, and such confined monopoles form either necklaces made of $N,2N,$ etc. monopoles, or mesons with a monopole and the corresponding antimonopole connected by two strings, see the review \cite{SYrev}.

Similarly, for the cases $r=1$ and $r=0$, confined quarks follow the same pattern.

\section{Further examples}

In this Appendix we provide a few more examples of the description of the confining picture, in addition to the example from Sec.~\ref{sec:example_U3_r1} above.

\subsection{Gauge group $U(5)$, the $r=2$ vacuum}
\label{U_5_r_2}

In this example, the UV gauge group is taken to be $U(5)$, and we consider the $r=2$ vacuum, where two of the quark flavors develop VEVs, cf. Eq.~\eqref{q}.

\subsubsection{$U(1)^5 \to U(1)_\text{unbr}$ effective theory}

Let us start by specifying a basis within the Cartan subalgebra $\mathfrak{u} (5)$
\begin{align}
	&H_1=\text{diag}(a,0,0,0,0), \quad  H_2=\text{diag}(0,a,0,0,0), \nonumber\\
	&H_{34}=\text{diag}(0,0,b,-b,0), \quad H_{45}=\text{diag}(0,0,f,f,-2f),\nonumber \\
	& H_{\text{unbr}}= \text{diag}(0,0,d,d,d); 
	\label{u5}
\end{align}
For the correct normalization in Eq.~\eqref{algebra_normalization}, the parameters here need to be chosen as $a^2=1/2, \ b^2=1/4, \ d^2=1/6,\ f^2=1/12$.
However, for now we will keep them arbitrary.

Our notation for the set (``vector'') of electric and magnetic charges under $U(1)^{5}$ is as follows:
\begin{align}
	&\vec{n}_{q^{kA}} =\left(\mathtt{Q}^{\text{el}}_{1},\mathtt{Q}^{\text{el}}_{2},\mathtt{Q}^{\text{el}}_{34},\mathtt{Q}^{\text{el}}_{45},\mathtt{Q}^{\text{el}}_{\text{unbr}} \right), \\
	&\vec{m}_{M_{ij}}=\left(\mathtt{Q}^{\text{mag}}_{1},\mathtt{Q}^{\text{mag}}_{2},\mathtt{Q}^{\text{mag}}_{34},\mathtt{Q}^{\text{mag}}_{45},\mathtt{Q}^{\text{mag}}_{\text{unbr}} \right).
\end{align}  

Again, the monopole charges are determined by the Cartan subalgebra of $\mathfrak{su}(5) \in \mathfrak{u}(5)$.
It is convenient to parametrize it by the generators \( H_{34} \) and \( H_{45} \) from \eqref{u5}, plus two more generators, \( H_{12} \) and \( H_{23} \), defined as
\begin{align}
	H_{12}= \text{diag}(-3x,0,x,x,x), \quad H_{23}=\text{diag}(y,-4y,y,y,y),
\end{align}
These generators have correct normalizations provided $x^2=1/24, \ y^2=1/40$. 
With these conventions, the simple roots of the \( \mathfrak{su}(5) \) subalgebra become
\begin{align}
	&\gamma_1=(-3x,5y,0,0), \quad \gamma_2=(-x,-5y,-b,-f) \nonumber \\
	&\gamma_{3}=(0,0,2b,0), \quad \gamma_4 = (0,0,-b,-3f),
\end{align}
This allows us to write down the magnetic charges of monopoles,
\begin{align}
	&\vec{m}_{M_{12}}=\left(a,-a,0,0,0 \right), \quad \vec{m}_{M_{23}}=\left(0,a,-b,-f,-d \right), \nonumber \\
	&\vec{m}_{M_{34}}=\left(0,0,2b,0,0 \right), \quad\vec{m}_{M_{45}}=\left(0,0,-b,-3f,0 \right).
	\label{example_U5_charges_monopoles}
\end{align}
The electric charges of the quarks are read off directly from Eq.~\eqref{u5}
\begin{align}
	&\vec{n}_{q^{1A}} = \left(a,0,0,0,0\right), \quad \vec{n}_{q^{2A}} = \left(0,a,0,0,0\right), \nonumber \\
	&\vec{n}_{q^{3A}} = \left(0,0,b,f,d\right), \quad \vec{n}_{q^{4A}} = \left(0,0,-b,f,d\right), \nonumber \\
	&\vec{n}_{q^{5A}} = \left(0,0,0,-2f,d\right).
	\label{example_U5_charges_quarks}
\end{align}
Note that the charges in Eqs.~\eqref{example_U5_charges_monopoles} and \eqref{example_U5_charges_quarks} do satisfy the Dirac quantization condition in Eq.~\eqref{dirak_quant}.

The UV gauge group is broken down as $U(5) \to U(1)^5$ at intermediate energies.
At low energies, the VEVs of the squarks $q^{11},q^{22}$ and the monopoles $M_{34},M_{45}$ break it further down to a single $U_\text{unbr}$.
Consequently, in this vacuum, one can consider two kinds of magnetic strings and another two kinds of electric strings, to which we now turn our attention.

\subsubsection{Electric and magnetic strings}

\begin{figure}[t!]
	\centering
	\includegraphics[width=1\textwidth]{
		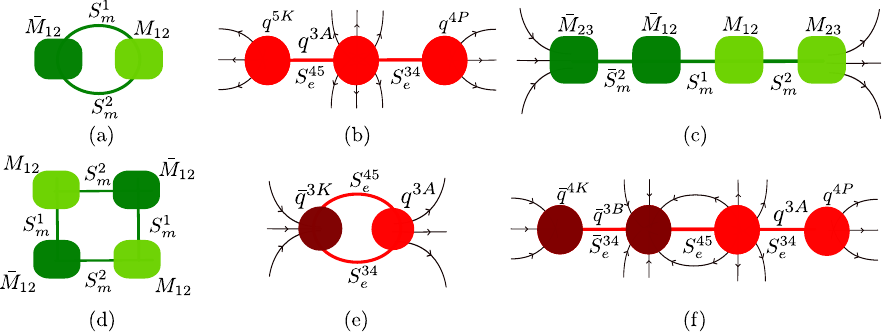} 
	\caption{Examples of states formed by the confined constituents (quarks and monopoles). The set of configurations includes both dipoles and necklaces consisting of several different charges. Junction-like quarks emit $U(1)_{\text{unbr}}$ in contrast to junction-like monopoles.}
	\label{fig:zoo}
\end{figure}

Each of the two condensed squarks is charged under a single  Abelian field
\begin{align}
	q^{11}: \ a A_{\mu}^{1}, \quad  q^{22}: \ a A_{\mu}^{2},
\end{align}
Correspondingly, the fluxes carried by the two magnetic strings are found as
\begin{align}
	\vec{S}_{m}^{1}=\left(\frac{1}{2a},0,0,0,0  \right), \quad \vec{S}_{m}^2=(0,\frac{1}{2a},0,0,0),
\end{align}
see \eqref{def_string_charge}.

These magnetic strings can confine those monopoles that did not form condensates.
As in the previous example, let us expand the charges of the confined monopoles $M_{12},M_{23}$ into confined $\vec{S}_{m}^1, \ \vec{S}_{m}^2$, screened $\vec{m}_{34},\vec{m}_{45}$, and unbroken $\vec{m}^{\text{unbr}}$ parts
\begin{align}
	&\vec{m}_{M_{12}}=\vec{S}_{m}^{1} - \vec{S}_{m}^2, \nonumber \\
	&\vec{m}_{M_{23}}=\vec{S}_{m}^{2}-\frac{1}{3}\vec{m}_{M_{34}}+\frac{1}{3}\vec{m}_{M_{45}}-\frac{\sqrt{6}}{6}\vec{m}^{\text{unbr}}.
	\label{su5_r=2_magnetic_charges_decomposition}
\end{align}
From Eq.~\eqref{su5_r=2_magnetic_charges_decomposition} we  see that the monopole \( M_{23} \) is attached to a single string carrying the flux $\vec{S}_{m}^{2}$, while the monopole \( M_{12} \) needs to be attached to two strings (or to a composite string) carrying both magnetic fluxes.
In other words, \( M_{23} \) can serve as an endpoint of a magnetic string, while \( M_{12} \) represents a string junction \cite{SYmon,T}.

Now we repeat the same argument but with the monopole condensates and magnetic strings.
The condensed monopoles $M_{34},M_{45}$ interact with the following combinations of Abelian fields:
\begin{align}
	M_{34}: \ 2b A^{D 34}_{\mu}, \quad M_{45}: -b A_{\mu}^{D23}-3fA_{\mu}^{D45}. 
\end{align}
Correspondingly, we find the string solution with the winding of $M_{34}$ and $M_{45}$
\begin{align}
	M_{34}\sim e^{i \alpha(x)}\sqrt{\frac{\xi_3}{2}} \quad \text{and} \quad M_{45} \sim  \sqrt{\frac{\xi_4}{2}}; \quad |x|\to \infty,
\end{align}
To ensure finite string tension, one must impose the condition 
\begin{align}
	2b  A^{D 34}_{\mu} \sim \partial_j\alpha, \quad -b A^{D 34}_{\mu} -3fA^{D 45}_{\mu}\sim0,
\end{align}
which gives
\begin{align}
	A^{D 34}_{\mu}\sim\frac{\partial_{j}\alpha}{2b}, \quad A^{D 45}_{\mu}\sim -\frac{\partial_j\alpha}{6f},
\end{align}
so the charges of the string are 
\begin{align}
	\vec{S}_{e}^{34}=(0,0,-\frac{1}{4b},\frac{1}{12f},0).
\end{align}
The charges of the second string are determined in a completely analogous manner:
\begin{align}
	\vec{S}_{e}^{45}=(0,0,0,\frac{1}{6f},0),
\end{align}
and eventually we obtain  
\begin{align}
	&\vec{n}_{q^{4A}}= \vec{S}_{e}^{34}+ \frac{1}{\sqrt{6}}\vec{n}^{\text{unbr}} \nonumber \\
	& \vec{n}_{q^{5A}}=- \vec{S}_{e}^{45}+ \frac{1}{\sqrt{6}}\vec{n}^{\text{unbr}} \nonumber \\
	&\vec{n}_{q^{3A}}= \vec{S}_{e}^{45}-\vec{S}_{e}^{34} +\frac{1}{\sqrt{6}}\vec{n}^{\text{unbr}}.
\end{align}

The picture of confined quarks differs slightly from the one with confined monopoles. 
The quarks \( q^{4A} \) and \( q^{5A} \) are attached to the endpoints of distinct strings, and both serve as sources of positive charge with respect to the massless \( U(1)_{\text{unbr}} \) field. There is also a single string junction --- the quark \( q^{3A} \) --- which, however, also carries a charge under the unbroken generator. The zoo of possible states is depicted in Fig.~\ref{fig:zoo}.

\subsection{Masses of confined constituents for $U(3),\, N_f=3$}
\label{append:n3}

\begin{figure}[t]
	\centering
	\includegraphics[width=1\textwidth]{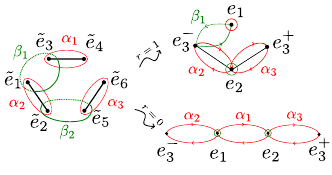} 
	\caption{$\alpha$ and $\beta$ contours for U(3). On the left, we show a generic point on the Coulomb branch, where the roots are denoted as $\Tilde{e}_k$. At a specific point, corresponding to the $r=1,\,r=0$ vacua of the $\mu$-deformed theory, some of the roots collide; at this point, the roots are denoted by untilded $e_k$, as in most of this paper.  The contours $\beta_1$ and $\beta_2$ determine the masses of the elementary monopoles $M_{12}$ and $M_{23}$, respectively, whereas the mass of the composite monopole $M_{13}$ is determined by the combined cycle $\beta_1+\beta_2$. The remaining three contours $\alpha_k$ determine the masses of the quark states $q^{kA}$.  }
	\label{fig:roots_append}
\end{figure}

We now turn to the explicit computation of the quark and monopole masses using the Seiberg--Witten solution for the $U(3)$ theory, considering the two cases with $r=0,1$. 

At a generic point on the Coulomb branch, the Seiberg--Witten curve is unfactorized and takes the following form: $ y^2=\prod_{k=1}^{6}(x-\Tilde{e}_k)$. One can relate the masses of the elementary monopole and quark hypermultiplets to homology cycles on the Seiberg--Witten curve as shown in Fig.~\ref{fig:roots_append}. 

\subsubsection{The $r=1$ vacuum}

In the $r=1$ vacuum, where the squark $q^{11}$ and the monopole $M_{23}$ condense, some roots collide, and the curve takes the form given in Eq.~\eqref{SW_curve_rep-01}.
Consequently, the contours $\alpha_1$ and $\beta_2$ shrink; see Fig.~\ref{fig:roots_append}. 
The remaining contours determine the masses of the confined states. 
Using the general result from Eqs.~\eqref{eP0} and \eqref{epm}, we find:
\begin{align}
	\sqrt{2}e_1\approx-m_1+\Lambda_{4D}^3 \frac{(m_2-m_3)(m_3-m_1)}{m_1^5}, \quad e_2\approx0, \quad \sqrt{2}e_3^{\pm}\approx\pm 2\sqrt[4]{\Lambda_{4D}^3\frac{m_2m_3}{m_1}}. 
	\label{roots_U(3)}
\end{align}

Let us focus on the monopole vacuum, so that the gluino condensate is given by 
\begin{align}
	\frac{S}{\mu} \approx\sqrt[2]{\Lambda_{4D}^3\frac{m_2m_3}{m_1}}=\Lambda_{YM}^2,
\end{align}
which coincides with the 2D result in Eq.~\eqref{t_eq}.
The exact formulas from Sec.~\ref{sec:masses_of_states}, together with the quasiclassical (large-$m$) approximation for the roots \eqref{roots_U(3)}, lead to the monopole masses
\begin{align}
	&M^{\text{monopole}}_{12}=M^{\text{monopole}}_{13}\approx \nonumber \\
	&\frac{1}{2 \pi} \Bigg|3 m_1 \log\left(\frac{m_1}{\Lambda_{4D}}\right)+ \sum_{A=2,3}\left[(m_A-m_1)\log\left(\frac{m_A-m_1}{m_1}\right)- m_A\log\left( \frac{m_A}{m_1}\right)\right]\Bigg| \,,
	\label{4D_mass_formula_monopoles_N=3}
\end{align}
and the quark masses
\begin{equation}
	m^{q}_{2,A}\approx\left|\frac{4}{\pi}\Lambda_{YM}+ m_A\right|,  \qquad
	m^{q}_{3,A}\approx\left|-\frac{4}{\pi}\Lambda_{YM}+ m_A\right|,  \quad A=1,2,3;
	\label{4D_mass_formula_quarks_N=3}
\end{equation}
One can immediately see that Eqs.~\eqref{4D_mass_formula_monopoles_N=3} and \eqref{4D_mass_formula_quarks_N=3} coincide with the corresponding 2D results from Eqs.~\eqref{m-kink_mass} and \eqref{q-kink_mass}.

\subsubsection{The $r=0$ vacuum}

In the case of $r=0$, all the $\beta$-cycles shrink to zero size, and the monopoles $M_{12}$ and $M_{23}$ condense. 
In the corresponding ``strand of beads'' configuration (analogous to Fig.~\ref{fig:zoo}(b)), the state $q^{1A}$ is a junction, whereas $q^{2A}$ and $q^{3A}$ are the endpoints. 
Using the 2D result \eqref{small_sigma} and the 2D--4D correspondence, we obtain the monopole roots 
\begin{equation}
	\sqrt{2}e^{\pm}_{3}\approx\pm2\Lambda_{YM}, \ \sqrt{2}e_{2}\approx\Lambda_{YM}, \ \sqrt{2}e_1\approx-\Lambda_{YM},
\end{equation}
where we defined
\begin{align}
	\Lambda_{YM}^2 \approx\Lambda_{4D}\sqrt[3]{m_1m_2m_3}.
\end{align}
Therefore, confined quarks have the following masses:
\begin{equation}
	\begin{aligned}
		m^{q}_{1,A}&= M^{kink}_{\sigma_2,\sigma_1}\approx|m_A+0|, \\ m^{q}_{3,A}&=M^{kink}_{\sigma^{+}_{3},\sigma_2}\approx|m_A-\frac{3\sqrt{3}\Lambda_{YM}}{\pi}|, \\ m^{q}_{2,A}&=M^{kink}_{\sigma_{1},\sigma^{-}_{3}}\approx|m_A+\frac{3\sqrt{3}\Lambda_{YM}}{\pi}|, \\
		A&=1,2,3.
	\end{aligned}
	\label{U3_r0_quark_masses}
\end{equation}

Let us discuss the physical implications of this result.
In Eq.~\eqref{U3_r0_quark_masses}, the confined quarks corresponding to different colors (but with the same fixed flavor $A$) have different masses.
This is unlike the situation in the ``real-world'' QCD.
This happens because the gauge group $U(3)$ here is broken down to $U(1)^3$ by the non-perturbative adjoint scalar condensation, where different components of the adjoint VEVs $\sim \Lambda_{YM}$ generally have different phases.
This aligns with the common lore that confinement in the SW theory is not quite the same as in QCD, as already follows from the sine law \eqref{sin_eq}.

All in all, here we have demonstrated that the ``extra'' splitting, known for the string tensions in the SW theory, also extends to the splitting of the masses of  the confined  quarks.

Let us  stress,  however, that all physical states are colorless. They can be charged only under the unbroken $U(1)_{\text{unbr}}$ group. The color  fluxes of quarks are absorbed by confining strings, while the Cartan ``remnants'' are screened by the adjoint condensate.

\section{Technicalities of the Seiberg-Witten solution }
\label{sec:appendix_SW_technical}

In this Appendix we provide some technical details related, e.g., to the computation of the periods of the SW curve.

\subsection{Roots of the Seiberg-Witten curve and the gaugino VEV  }
\label{append:roots_and_vevs}
We begin by evaluating the values of the roots of the SW curve and the values of the condensates for the case $r=N-2$. The Seiberg–Witten curve reads
\begin{align}
	y^2=\prod_{P=1}^N\left(x-\phi_P\right)^2-4\left(\frac{\Lambda_{4D}}{\sqrt{2}}\right)^{2 N-N_f} \prod_{A=1}^{N_f}\left(x+\frac{m_A}{\sqrt{2}}\right).
\end{align}
When $N-2$ quarks and one monopole condense, double roots appear as follows:
\begin{align}
	y^2=\prod_{P=1}^{N-2}(x-e_P)^2(x-e_{N-1})^2(x-e^{+}_N)(x-e^-_N),
\end{align}
where the double root $e_{N-1}$ corresponds to the condensed monopole.
Let us find corrections to \eqref{quasiroots}. Consider $x \propto \phi_{P_0} \propto -\frac{1}{\sqrt{2}}m_{P_0}$ for some $1\leq P_0\leq N-2$. Other values of $\phi_{P}$ will be taken to be classical.
\begin{align}
	y^2\approx \left(x-\phi_{P_{0}}\right)^2\prod_{P\neq P_0}^{N-2}&\left(-\frac{m_{P_0}}{\sqrt{2}}+\frac{m_{P}}{\sqrt{2}}\right)^2\frac{m_{P_0}^4}{4} \nonumber \\
	&-4\left (x+\frac{m_{P_0}}{\sqrt{2}}\right)\left(\frac{\Lambda_{4D}}{\sqrt{2}}\right)^{2 N-N_f} \prod_{A\neq P_0}^{N_f}\left(\frac{-m_{P_0}}{\sqrt{2}}+\frac{m_A}{\sqrt{2}}\right), 
\end{align}
Requiring the degeneracy of the roots, we obtain
\begin{align}
	& e_{P_0}\approx-\frac{m_{P_0}}{\sqrt{2}}+\frac{(\Lambda_{\mathcal{N}-2})^{2N-N_f}}{\sqrt{2} m_{P_0}^4}\frac{\prod_{\substack{ A=N-1} }^{N_f}\left(m_A -m_{P_0} \right)}{\prod_{\substack{ A=1} }^{N-2}\left(m_A -m_{P_0} \right)}, \ A\neq P_0 \label{eP0} \\
	& \phi_{P_0}\approx- \frac{m_{P_0}}{\sqrt{2}}-\frac{(\Lambda_{\mathcal{N}-2})^{2N-N_f}}{\sqrt{2} m_{P_0}^4}\frac{\prod_{\substack{ A=N-1} }^{N_f}\left(m_A -m_{P_0} \right)}{\prod_{\substack{ A=1} }^{N-2}\left(m_A -m_{P_0} \right)}, \ A\neq P_0
\end{align}
For small roots, consider $x \ll m_{P}$
\begin{align*}
	y^2\approx (x-\phi_{N-1})^2(x-\phi_N)^2\prod_{P=1}^{N-2}\frac{m_{P}^2}{2}-4 \left(\frac{\Lambda_{4D}}{\sqrt{2}}\right)^{2 N-N_f} \prod_{A=1}^{N_f}\frac{m_A}{\sqrt{2}}, 
\end{align*}
from which we find two options 
\begin{align}
	&\text{Monopole vacuum} & &\text{Dyon vacuum}  \nonumber \\
	&e_{N-1}\approx0  & & e_{N-1}\approx0  \nonumber \\
	&e_N^{\pm}\approx \pm \sqrt[4] {4\Lambda_{4D}^{2N-N_f}\frac{\prod_{A=N-1}^{N_f}m_A}{\prod_{P=1}^{N-2}m_{P}}}=\pm \sqrt{2}\Lambda_{YM} & &e_N^{\pm}\approx \pm i \sqrt[4] {4\Lambda_{4D}^{2N-N_f}\frac{\prod_{A=N-1}^{N_f}m_A}{\prod_{P=1}^{N-2}m_{P}}}=\pm i\sqrt{2}\Lambda_{YM} \nonumber \\
	&\phi_{N}\approx-{\phi}_{N-1}\approx  \sqrt[4]{\Lambda_{4D}^{2N-N_f}\frac{\prod_{A=N-1}^{N_f}m_A}{\prod_{P=1}^{N-2}m_{P}}} & &\phi_{N}\approx-{\phi}_{N-1}\approx i \sqrt[4]{\Lambda_{4D}^{2N-N_f}\frac{\prod_{A=N-1}^{N_f}m_A}{\prod_{P=1}^{N-2}m_{P}}} \label{epm}
\end{align}
The gluino condensate is 
\begin{align}
	\frac{S}{\mu}= \frac{(e_{N}^{\pm})^2}{2}\approx\pm\ \sqrt{\Lambda^{2N-N_f}\frac{\prod_{A=N-1}^{N_f}m_A}{\prod_{P=1}^{N-2}m_{P}}}= \pm \Lambda_{YM}^2,
\end{align}
where the plus sign corresponds to the monopole vacuum and the minus sign to the dyon one. 

The results above confirm our conclusion in Eq.~\eqref{equalroots_r} that the 2D  VEVs of $\sigma$ coincide with the roots of the Seiberg--Witten curve for the $r=N-2$ vacuum in the large-mass limit; see \eqref{large_sigma} and \eqref{small_sigma}. In particular, the root $e_{N-1}\approx 0$ corresponds 
to the VEV $\sigma_{k=s+1}\approx 0$ in \eqref{small_sigma} for $r=N-2$.

\subsection{Resolvent in the hybrid $r$-vacuum}
\label{resolvent_in_r}
Now we prove formula \eqref{exact_resolvent}:
\begin{align}
	\left\langle\Tr\frac{1}{\Sigma-\sqrt{2}\Phi}\right\rangle  = &\frac{1}{2} \sum_{A=1}^{N_f}\frac{1}{\Sigma+m_A}+\frac{1}{2}\frac{2N-N_f}{\sqrt{\Sigma^2- \frac{4 S}{\mu}}}\nonumber\\ -&\frac{1}{2}\sum^r_{A=1} \frac{\sqrt{m_A^2-\frac{4S}{\mu}}}{\sqrt{\Sigma^2-\frac{4S}{\mu}}(\Sigma+m_A)}+\frac{1}{2}\sum^{N_f}_{A=r+1} \frac{\sqrt{m_A^2-\frac{4S}{\mu}}}{\sqrt{\Sigma^2-\frac{4S}{\mu}}(\Sigma+m_A)}.
	\label{exact_resolvent_app}
\end{align}
Recall that the resolvent is related to the Seiberg--Witten differential by 
\begin{align}
	x\left\langle \Tr\frac{1}{x-\Phi} \right\rangle dx=\lambda_{\mathrm{SW}}=\frac{x dP}{y}-\frac{xP}{2 y}\frac{dQ}{Q}+\frac{x}{2}\frac{dQ}{Q}, \label{c9}
\end{align}
where
\begin{align}
	& P=\prod_{k=1}^{N}\left(x-\phi_{k}\right)  , \ Q = \prod_{P=1}^{N_f}\left(x+\frac{m_P}{\sqrt{2}}\right), \ y^2 = (x^2-e_N^2)\prod_{k=1}^{N-1}(x-e_k)^2.
\end{align}
First, we separate the poles of $\lambda_{\mathrm{SW}}$
\begin{align}
	& \frac{x P'}{\sqrt{x^2-e_{N}^2}\prod_{k=1}^{N-1}(x-e_{k})}-\frac{xP}{2 \sqrt{x^2-e_{N}^2}\prod_{k=1}^{N-1}(x-e_{k})}\frac{Q'}{Q}+\frac{x}{2}\frac{Q'}{Q} \nonumber \\
	&=\sum_{k=1}^{N-1}\frac{G(x)-\sum_{P=1}^{N_f}\frac{F(x)}{2\left(e_{k}+\frac{m_{P}}{\sqrt{2}}\right)}}{(x-e_k)\prod^{N-1}_{s \neq k}(e_k-e_s)}+\sum_{P=1}^{N_f}\frac{1}{x+\frac{m_{p}}{\sqrt{2}}}\left(\frac{x}{2}+\frac{(-1)^NF(x)}{2\prod_{k=1}^{N-1}\left(\frac{m_{j}}{\sqrt{2}}+e_{k}\right)} \right),
\end{align}
where we used the notation 
\begin{align}
	G(x) = \frac{xP'}{\sqrt{x^2-e_{N}^2}}, \quad F(x)= \frac{x P}{\sqrt{x^2-e_{N}^2}}.
\end{align}
The analytical properties of the differential require 
\begin{align}
	&G(e_k)-\sum_{P=1}^{N_f}\frac{F(e_k)}{2\left(e_{k}+\frac{m_{P}}{\sqrt{2}}\right)}=0, \quad k=1,\ldots,N-1 ; \label{cond1} \\
	&-\frac{m_{P}}{2\sqrt{2}}+\frac{(-1)^N F(-\frac{m_P}{\sqrt{2}})}{2\prod_{k=1}^{N-1}\left(\frac{m_{j}}{\sqrt{2}}+e_{k}\right)}=n_P\frac{m_{P}}{\sqrt{2}}, \quad P=1,\ldots N_f. \label{cond2}
\end{align}
The first condition ensures analyticity with respect to the roots. The second one determines whether the quark is massless on the Coulomb branch or not; see  \eqref{res_masses}, so we should choose $n_{P}=-1$ for massless quarks $P=1,\ldots,r$ and $n_{P}=0$ for $P=r+1,\ldots,N_f$. From \eqref{cond1} and \eqref{cond2}, we obtain 
\begin{align}
	& P'(e_k)= \sum_{P=1}^{N_f}\frac{P(e_k)}{2(e_k+\frac{m_P}{\sqrt{2}})} ,\quad k=1,\ldots,N-1 ; \label{cond11}\\
	&\frac{(-1)^NP(-\frac{m_{P}}{\sqrt{2}})}{\prod_{k=1}^{N-1}(\frac{m_{P}}{\sqrt{2}}+e_{k})} = - \frac{2n_P+1}{\sqrt{2}}\sqrt{m_{P}^2-2e_{N}^2}, \quad P=1,\ldots,N_f. \label{cond22}
\end{align}
This will allow us to eliminate the $\phi_k$ values. Let us start from the right-hand side of equation \eqref{c9}
\begin{align}
	&\frac{x dP}{y}- \frac{xPdQ}{2yQ}+\frac{xdQ}{2Q}  = \\
	&\frac{xdx}{\sqrt{x^2-e_{N}^2}}\left[\frac{P'}{\prod_{k=1}^{N-1}(x-e_k)}-\sum_{P=1}^{N_f}\frac{P}{2\prod_{k=1}^{N-1} (x-e_k)}\frac{1}{x+\frac{m_{P}}{\sqrt{2}}}\right]+ \frac{xdQ}{2Q} \label{b16}.
\end{align}
Consider the second term in square brackets. One can again separate the poles and then use \eqref{cond11}, \eqref{cond22} 
\begin{align}
	&\sum_{P=1}^{N_f}\frac{P}{2\prod_{k=1}^{N-1} (x-e_k)}\frac{1}{x+\frac{m_{P}}{\sqrt{2}}}  \nonumber\\
	&=\frac{N_f}{2}+\sum_{k=1}^{N-1}\frac{P'(e_k)}{\prod_{s\neq k}(x_k-e_s)(x-e_k)}-\frac{1}{2}\sum_{P=1}^{N_f}\frac{(2n_{P}+1)}{\sqrt{2}}\frac{\sqrt{m_{P}^2-2e_{N}^2}}{x+\frac{m_{P}}{\sqrt{2}}}.\label{b17}
\end{align}
Working with the first term in square brackets of formula \eqref{b16}, we find 
\begin{align}
	\frac{P'}{\prod_{k=1}^{N-1}(x-e_k)}=N+\sum_{k=1}^{N-1}\frac{P'(e_k)}{\prod_{s\neq k}(x_k-e_s)(x-e_k)} \label{b18}.
\end{align}
Substitution of \eqref{b17} and \eqref{b18} back into \eqref{b16} leads us to 
\begin{align}
	\lambda_{\text{SW}} =  \frac{xdx}{2\sqrt{x^2-e_{N}^2}}\left[ 2N-N_f+\sum_{P=1}^{N_f}\frac{(2n_{P}+1)\sqrt{m_P^2-2e_{N}^2}}{\sqrt{2}x+m_{P}}\right]+\frac{xdQ}{2Q}.
\end{align}
Changing variables   $\sqrt{2}x=\Sigma$, substituting  $2e^{2}_{N}=\frac{4S}{\mu}$, and comparing it with \eqref{c9}, we eventually arrive at \eqref{exact_resolvent_app}.


\newpage
\addcontentsline{toc}{section}{References}
\begin{thebibliography}{99}



\bibitem{SW}
N.~Seiberg and E.~Witten,
{\em Electric - magnetic duality, monopole condensation, and confinement in N=2 supersymmetric Yang-Mills theory,}
Nucl. Phys. B \textbf{426} (1994), 19-52
[erratum: Nucl. Phys. B \textbf{430} (1994), 485-486]
[arXiv:hep-th/9407087 [hep-th]].





\bibitem{SW2}
N.~Seiberg and E.~Witten,
{\em Monopoles, duality and chiral symmetry breaking in N=2 supersymmetric QCD,}
Nucl. Phys. B \textbf{431} (1994), 484-550
[arXiv:hep-th/9408099 [hep-th]].



\bibitem{HT1}
A.~Hanany and D.~Tong,
{\em Vortices, instantons and branes,}
JHEP {\bf 0307}, 037 (2003).
[hep-th/0306150].


\bibitem{ABEKY}
R.~Auzzi, S.~Bolognesi, J.~Evslin, K.~Konishi and A.~Yung,
{\em Non-Abelian superconductors: Vortices and
confinement in ${\mathcal N}=2$  SQCD,}
Nucl.\ Phys.\ B {\bf 673}, 187 (2003).
[hep-th/0307287].


\bibitem{SYmon}
M.~Shifman and A.~Yung,
{\em Non-Abelian string junctions as confined monopoles,}
Phys.\ Rev.\ D {\bf 70}, 045004 (2004)
[hep-th/0403149].


\bibitem{HT2}
A. Hanany and D. Tong,
{\em Vortex strings and four-dimensional gauge dynamics,}
JHEP {\bf 0404}, 066 (2004)
[hep-th/0403158].


\bibitem{Trev}
D.~Tong, {\em TASI Lectures on Solitons,}
arXiv:hep-th/0509216.


\bibitem{Jrev}
M.~Eto, Y.~Isozumi, M.~Nitta, K.~Ohashi and N.~Sakai,
{\em Solitons in the Higgs phase: The moduli matrix approach,}
J.\ Phys.\ A  {\bf 39}, R315 (2006)
[arXiv:hep-th/0602170].


\bibitem{SYrev}
M. Shifman and A. Yung,
{\em Supersymmetric Solitons and How They Help Us Understand Non-Abelian Gauge Theories},
Rev.\ Mod.\ Phys.\  {\bf 79}, 1139 (2007)
[hep-th/0703267]; for an expanded version see
{\em Supersymmetric Solitons,}
(Cambridge University Press, 2009).



\bibitem{Trev2}
D.~Tong,
{\em Quantum Vortex Strings: A Review,}
Annals Phys.\  {\bf 324}, 30 (2009)
[arXiv:0809.5060 [hep-th]].




\bibitem{FI}
P.~Fayet and J.~Iliopoulos,
{\em Spontaneously Broken Supergauge Symmetries and Goldstone Spinors,}
Phys.\ Lett.\  B {\bf 51}, 461 (1974).



\bibitem{Vachaspati:1991dz}
T.~Vachaspati and A.~Achucarro,
{\em Semilocal cosmic strings,}
Phys. Rev. D \textbf{44}, 3067-3071 (1991)
doi:10.1103/PhysRevD.44.3067



\bibitem{AchVas}
For a review see e.g. A.~Achucarro and T.~Vachaspati,
{\em Semilocal and electroweak strings,}
Phys.\ Rept.\  {\bf 327}, 347 (2000)
[hep-ph/9904229].


\bibitem{SYsem}
M.~Shifman and A.~Yung,
{\em Non-Abelian semilocal strings in  ${\mathcal N} = 2$ supersymmetric QCD,}
Phys.\ Rev.\  D {\bf 73}, 125012 (2006)
[arXiv:hep-th/0603134].


\bibitem{Jsem}
M.~Eto, J.~Evslin, K.~Konishi, G.~Marmorini, et al.,
{\em On the moduli space of semilocal strings and lumps,}
Phys.\ Rev.\  D {\bf 76}, 105002 (2007)
[arXiv:0704.2218 [hep-th]].

    
    
\bibitem{SVY}
M.~Shifman, W.~Vinci and A.~Yung,
{\em Effective World-Sheet Theory for Non-Abelian Semilocal 
Strings in ${\mathcal N} = 2$ Supersymmetric QCD,}
Phys.\ Rev.\ D {\bf 83}, 125017 (2011)
[arXiv:1104.2077 [hep-th]].


\bibitem{Dorey}
N.~Dorey,
{\em The BPS spectra of two-dimensional supersymmetric gauge theories with  twisted mass terms,}
JHEP {\bf 9811}, 005 (1998) [hep-th/9806056].


\bibitem{DoHoTo}
N.~Dorey, T.~J.~Hollowood and D.~Tong,
{\em The BPS spectra of gauge theories in two and four dimensions,}
JHEP {\bf 9905}, 006 (1999)
[arXiv:hep-th/9902134].



\bibitem{Shifman:2014lba}
M.~Shifman and A.~Yung,
{\em Quantum Deformation of the Effective Theory on Non-Abelian string and 2D-4D correspondence,}
Phys. Rev. D \textbf{89} (2014) no.6, 065035
[arXiv:1401.1455 [hep-th]].



\bibitem{Gaiotto:2013sma}
D.~Gaiotto, S.~Gukov and N.~Seiberg,
{\em Surface Defects and Resolvents,}
JHEP \textbf{09} (2013), 070
[arXiv:1307.2578 [hep-th]].





\bibitem{ANO}
A.~A.~Abrikosov,
{\em On the Magnetic Properties of Superconductors of the Second Group,}
Sov.~Phys.~JETP {\bf 5}, no 6, 1174 (1957)
[Zh.~Eksp.~Teor.~Fiz. {\bf 32}, no 6, 1442 (1957)];
%
H.~B.~Nielsen and P.~Olesen,
{\em Vortex-Line Models for Dual Strings,}
Nucl.~Phys.~B {\bf 61}, 45 (1973).
[Reprinted in {\em Solitons and Particles}, Eds. C.~Rebbi and G.~Soliani
(World Scientific, Singapore, 1984), p.~365].

\bibitem{HanStrassZaf}
A.~Hanany, M.~J.~Strassler and A.~Zaffaroni,
{\em Confinement and strings in M{QCD}},
Nucl.\ Phys.\ B {\bf 513}, 87 (1998)
[hep-th/9707244].


\bibitem{VY}
A.~I.~Vainshtein and A.~Yung,
{\em Type I superconductivity upon
monopole condensation in Seiberg--Witten  theory,}
Nucl.\ Phys.\ B {\bf 614}, 3 (2001)
[hep-th/0012250].

\bibitem{Argyres:1996eh}
P.~C.~Argyres, M.~R.~Plesser and N.~Seiberg,
{\em The Moduli space of vacua of N=2 SUSY QCD and duality in N=1 SUSY QCD,}
Nucl. Phys. B \textbf{471}, 159-194 (1996)
[arXiv:hep-th/9603042 [hep-th]].




\bibitem{HaOz}
A.~Hanany and  Y.~Oz,
{\em On the Quantum Moduli Space of Vacua of $N=2$ Supersymmetric $SU(N_c)$ Gauge Theories,}
Nucl. \ Phys. \ B {\bf 452}, 283 (1995)
[hep-th/9505075].




\bibitem{Cachazo:2001jy}
F.~Cachazo, K.~A.~Intriligator and C.~Vafa,
{\em A Large N duality via a geometric transition,}
Nucl. Phys. B \textbf{603}, 3-41 (2001)
[arXiv:hep-th/0103067 [hep-th]].



\bibitem{Balasubramanian:2003tv}
V.~Balasubramanian, B.~Feng, M.~x.~Huang and A.~Naqvi,
{\em Phases of N=1 supersymmetric gauge theories with flavors,}
Annals Phys. \textbf{310}, 375-427 (2004)
[arXiv:hep-th/0303065 [hep-th]].



\bibitem{Shifman:2013zsa}
M.~Shifman and A.~Yung,
{\em Hybrid $r$-Vacua in $\mathcal{N}=2$ Supersymmetric QCD: Universal Condensate Formula,}
Phys. Rev. D \textbf{87}, no.8, 085044 (2013)
[arXiv:1303.1449 [hep-th]].




\bibitem{W79} 
E.~Witten,
{\em Instantons, The Quark Model, And The 1/N Expansion,}
Nucl.\ Phys.\ B {\bf 149}, 285 (1979).


\bibitem{W93}
E.~Witten,
{\em Phases of N = 2 theories in two dimensions,}
Nucl.\ Phys.\ B {\bf 403}, 159 (1993)
[hep-th/9301042].


\bibitem{Shifman:2010kr}
M.~Shifman and A.~Yung,
{\em Moduli Space Potentials for Heterotic non-Abelian Flux Tubes: Weak Deformation,}
Phys. Rev. D \textbf{82}, 066006 (2010)
[arXiv:1005.5264 [hep-th]].



\bibitem{Bolokhov:2013bea}
P.~A.~Bolokhov, M.~Shifman and A.~Yung,
{\em Twisted-Mass Potential on the Non-Abelian String World Sheet Induced by Bulk Masses,}
Phys. Rev. D \textbf{88}, 085016 (2013)
[arXiv:1308.4494 [hep-th]].




\bibitem{SYrvacua}
M.~Shifman and A.~Yung,
{\em r Duality and 'Instead-of-Confinement' Mechanism in \none   Supersymmetric QCD},
Phys.\ Rev.\  D {\bf 86}, 025001 (2012)
arXiv:1204.4165 [hep-th].




\bibitem{Veneziano:1982ah}
G.~Veneziano and S.~Yankielowicz,
{\em An Effective Lagrangian for the Pure N=1 Supersymmetric Yang-Mills Theory,}
Phys. Lett. B \textbf{113}, 231 (1982)




\bibitem{AdDVecSal}
A.~D'Adda, A.~C.~Davis, P.~DiVeccia and P.~Salamonson,
{\em An effective action for the supersymmetric CP$^{n-1}$ models,}
Nucl.\ Phys.\ {\bf B222} 45 (1983).



\bibitem{Cecotti:1992rm}
S.~Cecotti and C.~Vafa,
{\em On classification of N=2 supersymmetric theories,}
Commun. Math. Phys. \textbf{158}, 569-644 (1993)
[arXiv:hep-th/9211097 [hep-th]].




\bibitem{HaHo}
A.~Hanany and K.~Hori,
{\em Branes and N = 2 theories in two dimensions,}
Nucl.\ Phys.\  B {\bf 513}, 119 (1998)
[arXiv:hep-th/9707192].



\bibitem{Bolokhov:2012dv} 
P.~A.~Bolokhov, M.~Shifman and A.~Yung,
{\em 2D-4D Correspondence: Towers of Kinks versus Towers of Monopoles in N=2 Theories,}
Phys.\ Rev.\ D {\bf 85}, 085028 (2012)
[arXiv:1202.5612 [hep-th]].



\bibitem{Cachazo:2003yc}
F.~Cachazo, N.~Seiberg and E.~Witten,
{\em Chiral rings and phases of supersymmetric gauge theories,}
JHEP \textbf{04}, 018 (2003)
[arXiv:hep-th/0303207 [hep-th]].




\bibitem{DougShenk}
M.~R.~Douglas and S.~H.~Shenker,
{\em Dynamics of SU(N) supersymmetric gauge theory,}
Nucl.\ Phys.\ B {\bf 447}, 271 (1995)
[hep-th/9503163].

\bibitem{Strassler:1997ny}
M.~J.~Strassler,
{\em Messages for QCD from the superworld,}
Prog. Theor. Phys. Suppl. \textbf{131}, 439-458 (1998)
[arXiv:hep-lat/9803009 [hep-lat]].


\bibitem{GMMM}
A.Gorsky, A.Marshakov, A.Mironov and A.Morozov, 
{\em N=2 Supersymmetric QCD and Integrable Spin Chains: Rational Case $N_f < 2N_c$,}
Phys. \ Lett.\ {\bf B380}, 75 (1996) 
[hep-th/9603140]


\bibitem{T}
D.~Tong,
{\em Monopoles in the Higgs phase,}
Phys.\ Rev.\ D {\bf 69}, 065003 (2004)
[hep-th/0307302].


\end{thebibliography}
\end{document}